\documentclass[journal]{IEEEtran}
\renewcommand{\figurename}{Fig.}
\usepackage{cite}

\ifCLASSINFOpdf
\else
\fi
\ifCLASSOPTIONcompsoc
  \usepackage[caption=false,font=normalsize,labelfont=sf,textfont=sf]{subfig}
\else
  \usepackage[caption=false,font=footnotesize]{subfig}
\fi
\usepackage[T1]{fontenc}
\usepackage[utf8]{luainputenc}

\usepackage{afterpage}
\usepackage{float}
\usepackage{graphicx}
\usepackage{epstopdf}
\usepackage{epsfig}
\AppendGraphicsExtensions{.tif}

\usepackage{gensymb}
\usepackage{algorithm}
\usepackage{algorithmicx}
\usepackage{amsmath}
\usepackage{amssymb}

\usepackage{threeparttable}
\usepackage{array}
\usepackage{xcolor}
\usepackage{bm}
\usepackage{mathrsfs}

\begin{document}
%
\title{3D Markerless Tracking of Human Gait by Geometric Trilateration of Multiple Kinects}
%
%
%

\author{Lin Yang, Bowen Yang, Haiwei Dong,~\IEEEmembership{Senior Member,~IEEE}, and Abdulmotaleb El Saddik,~\IEEEmembership{Fellow,~IEEE}
\thanks{L. Yang, B. Yang, H. Dong, and A. El Saddik are with Multimedia Computing Research Laboratory (MCRLab), School of Electrical Engineering and Computer Science, University of Ottawa, 800 King Edward, Ottawa, ON, K1N 6N5, Canada (e-mail: \{lyang103; byang078; hdong; elsaddik\}@uottawa.ca).}
}
\maketitle


%

\begin{abstract}
In this paper, we develop an integrated markerless gait tracking system with three Kinect v2 sensors. A geometric principle-based trilateration method is proposed for optimizing the accuracy of the measured gait data. To tackle the data synchronization problem among the Kinect clients and the server, a synchronization mechanism based on NTP (Network Time Protocol) is designed for synchronizing the server and Kinect clients' clocks. Furthermore, a time schedule is designed for timing each Kinect client's data transmission. In the experiment, participants are asked to perform a 60 s walk while the proposed tracking system obtains the participant's gait data. Six joints (including left hip, right hip, left knee, right knee, left ankle and right ankle) of the participants are tracked where the obtained gait data are described as 6000 {movements} of joint positions (1000 {movements} for each joint). The results show that the trilateration tracking result by the three Kinect sensors has a much higher accuracy compared with the accuracy measured by a single Kinect sensor. Within a randomly sampled time period (67.726 s in the experiment), 98.37$\%$ of the frames generated by the gait tracking system have timing errors less than 1 ms, which is much better than the default NTP service embedded in the Windows 8.1 operating system. {The accuracy of the proposed system is  quantitatively evaluated and verified by a comparison with a commercial medical system (Delsys Trigno Smart Sensor System).}
\end{abstract}


%
\IEEEpeerreviewmaketitle

\section{Introduction}

\IEEEPARstart{I}{n} the late 1970s, the motion capture technique was first proposed by Johansson \cite{intro_mc_1} in a moving light display (MLD) experiment. Different from the older techniques applied in animation, such as rotoscoping or traditional animation, the movements of a human or nonhuman body part was recorded for the first time while ignoring the human or nonhuman body's visual appearance. The development of the motion capture technique was promoted in the 1980s by researchers such as Carol from MIT \cite{intro_mc_4}, who addressed the challenges in person-modeling 3D animation systems. In the 1990s, motion capture technologies continued to mature and new processes were developed, such as a CCD-camera-based system capable of biomechanical motion analysis \cite{intro_mc_5} and a visual motion estimation technology for recovering complex motions \cite{intro_mc_10}. In recent years, motion capture technologies have developed even more rapidly, and many motion capture systems have been promoted in the marketplace (such as {Optitrack}, Vicon motion capture system and Xsens Moven). They can generally be divided into several categories: optical systems, inertial systems, magnetic systems, mechanical systems, acoustic systems and markerless systems \cite{intro_mc_7,intro_mc_8,intro_mc_9}. The technical details will be discussed in Section \ref{sc:Related Work}.

By assessing human motion, motion capture technology helps people understand how human bodies perceive the world and benefits people by providing a more creative and healthier environment. Motion capture technology is gaining in popularity due to its high performance in entertainment, the military, biomechanics, and other areas. 

In the military, motion capture systems mainly focus on training for soldiers or simulating specific battle scenarios \cite{intro_mc_11}. As a small mistake might lead to a serious consequence and as live drills in the field are costly in terms of time and money, motion capture technology has potential value if relative real scenarios or soldiers' motions can be captured and visualized in real time. 

In {the} entertainment field, motion capture techniques are mainly applied in {Computer Graphic (CG)} video creation and game development. Typically, two methods are used to apply motion capture technology: 1) designing and recording specific movements or actions in reality with motion capture technology and mapping the movements or actions to the 3D models of the characters in games or CG video creation, which provides a vivid simulation of human motion and 2) capturing the player's motions, mapping the motions to the characters in the game, and showing how the players interact with the video games accordingly in real time (e.g., Kinect Sports). 

In biomechanics, motion capture technology is mainly applied for observing and recording the human body's biomechanical data. With these data, many applications for understanding human motion, improving the physical quality of the human body, and detecting disease have been developed. One of the most popular {biomechanical} applications with motion capture technology is sports analysis. {Experts of almost every kind of sport analyze} skill levels to determine how players can increase performance while decreasing the potential for injuries \cite{intro_mc_16,intro_mc_17,intro_mc_18}. Sports analysis applications apply motion capture technology to collect biomechanical data, analyze the data from a clinical perspective, define the user's skill level, observe potential injury causes, and report on how to exercise correctly or manipulate/adjust fitness equipments automatically (e.g., adjusting the speed of a treadmill, controlling the duration of running, and giving running posture advice). 

However, there are several common issues with the current motion capture systems.

\begin{itemize}
\item \textbf{\textit{{Economic Cost.}}} Usually, a motion capture system requires professional hardware devices for reaching high accuracy (e.g., for gait analysis, a Vicon system \cite{intro_mc_14} for gait tracking has a starting price from \$12,000, excluding training fees, calibration fees, installation fees, and so on) \cite{Intro_mc_19}.

\item \textbf{\textit{Wearable gear is required.}} Most motion capture technologies (except markerless motion capture technologies) require specific equipments to be worn on the human body (e.g., optical systems require markers to be attached to the body such that the specific cameras can detect the joints from a background), which usually limits the freedom of movement of the human body.

 \item \textbf{\textit{Complex setup.}} Different motion capture technologies are based on different locating approaches, which means that the requirements of the system environment are different. Usually, for motion capture, more equipments (more wearable gear, more observing cameras, etc.) lead to better accuracy and result in a more complex setup. For example, in a Vicon system, tens of cameras need to be positioned accordingly, and complex calibration procedures need to be conducted for better accuracy (this is why there are several costly courses for a Vicon system's installation, debugging, calibration, etc.).
\end{itemize}

Recently, Kinect for Windows v2 sensor \cite{Zhang_2015, Figueroa_2015} has been released with improved specifications. With its depth sensing and easy setup features, {the} development of a cost-effective markerless gait tracking system becomes possible. In this paper, we propose a trilateration method for improving Kinect sensor's accuracy and further develop an integrated gait tracking system based on the trilateration method. {This system consists of three Kinect clients obtaining and forwarding gait parameters and a server optimizing the gait parameters with the proposed trilateration method.} To tackle the data synchronization problem existing among the clients and the server, a synchronization mechanism based on NTP (Network Time Protocol) is then designed for synchronizing the server and Kinect clients' clocks, and a time schedule is designed for timing each Kinect client's data transmission. 

{Different from {\cite{SO_1}}, which proposed a trilateration method with 4 Kinects for solving differential equations, we further improve that method by applying geometric principles, using which only 3 Kinects rather than 4 are sufficient for obtaining the estimated position. In addition, to develop the whole system, we propose a synchronism scheme and a server-client structure for synchronizing the Kinect clients with the server with optimal transmitting performance. Finally, we conduct a thorough accuracy analysis of our proposed gait tracking system and a commercial medical tracking system, i.e., Delsys Trigno{\textsuperscript{TM}} Wireless Smart Sensor System, to evaluate the accuracy of our developed system.}

\section{\label{sc:Related Work}Related Work}

\subsection{Motion Capture Technologies}
At present, motion capture technologies can be classified into six categories: 

1) In optical motion capture technology, multiple high-speed cameras are positioned at fixed positions around the measurement area. Several markers that are able to reflect the special light beams from the cameras (or actively send special light beams to the cameras) are required to be attached on the joint positions. Thus, if one specific point can be observed by any two cameras, the {point's} position can be located in a 3D space based on the triangulation principle. By continuing to capture the frames with high-speed cameras, the points are tracked with post-data-processing procedures (e.g., 2D-to-3D visual human motion conversion) \cite{related_mc_1, RL_7}.
	\begin{table}
		\begin{centering}
			\protect\caption{\label{tab:mc_comparisions}Comparison of Motion Capture Technologies}
			\begin{tabular}{|>{\centering}p{1cm}|>{\raggedright}p{1.5cm}|>{\raggedright}p{2cm}|>{\raggedright}p{2.5cm}|}
				\hline 
				{\scriptsize{}Type} & {\scriptsize{}Device} & {\scriptsize{}Advantages} & {\scriptsize{}Disadvantages}\tabularnewline
				\hline 
				\hline 
				{\scriptsize{}Optical \cite{related_mc_1}} 
			  & {\scriptsize{}- Passive or Active marker}{\scriptsize \par}
			  
			    {\scriptsize{}- High-speed cameras}{\scriptsize \par}
			    
			     {\scriptsize{}-Data-processing device}{\scriptsize \par}
			  
			  & {\scriptsize{}- High accuracy}{\scriptsize \par}
				
				{\scriptsize{}- Relative high freedom of movement}{\scriptsize \par}
				
				{\scriptsize{}- No additional post-processing procedures (active markers)}{\scriptsize \par}
				
				{\scriptsize{}- Capturing high-speed motion} 
				
			  & {\scriptsize{}- Easily affected by surrounding light sources}{\scriptsize \par}
				
				{\scriptsize{}- Limited measurement range}{\scriptsize \par}
				
				{\scriptsize{}- Additional post-processing procedures (passive markers)}{\scriptsize \par}
				
				{\scriptsize{}- Additional wires or batteries are required to be worn for markers' power (active markers)}{\scriptsize \par}
				
				{\scriptsize{}- Complex setup and calibration}{\scriptsize \par}
				
				{\scriptsize{}- Costly}\tabularnewline
				\hline 
				{\scriptsize{}Inertial \cite{related_mc_2}} 
				
			  & {\scriptsize{}- MEMS sensors (accelerometers and gyroscopes)} {\scriptsize \par}
			  
			    {\scriptsize{}-Data-processing device}{\scriptsize \par}
			  
			  & {\scriptsize{}-No limitation for the measurement environment}{\scriptsize \par}
				
				{\scriptsize{}-No cameras are required}{\scriptsize \par}
				
				{\scriptsize{}- Large measurement area and portable}{\scriptsize \par}
				
				{\scriptsize{}- Real time} 
				
			  & {\scriptsize{}- Accuracy error accumulates with time}{\scriptsize \par}
				
				{\scriptsize{}- Accuracy may vary at different positions of the human body}\tabularnewline
				\hline 
				{\scriptsize{}Magnetic \cite{related_mc_4}} 
				
			  & {\scriptsize{}- Electromagnetic emission source and receivers}{\scriptsize \par}
			  
			    {\scriptsize{}- Data-processing device}{\scriptsize \par}
				
			  & {\scriptsize{}- Low cost}{\scriptsize \par}
				
				{\scriptsize{}- Mature technique}{\scriptsize \par}
				
				{\scriptsize{}- Real time} & {\scriptsize{}- Low accuracy}{\scriptsize \par}
				
				{\scriptsize{}- Limited size of measurement area}{\scriptsize \par}
				
				{\scriptsize{}- Transmission interference}{\scriptsize \par}
				
				{\scriptsize{}- No metallic object nearby}{\scriptsize \par}
				
				{\scriptsize{}- Low capability for capturing high-speed movements}\tabularnewline
				\hline 
				{\scriptsize{}Mechanical \cite{related_mc_6}} 
				
			  & {\scriptsize{}- Mechanical exoskeleton}{\scriptsize \par}
			    
			    {\scriptsize{}- Data-processing device}{\scriptsize \par}
				
			  & {\scriptsize{}- High efficiency}{\scriptsize \par}
				
				{\scriptsize{}- Real time}{\scriptsize \par}
				
				{\scriptsize{}- High accuracy}{\scriptsize \par}
				
				{\scriptsize{}- No limitation to the measurement environment} & {\scriptsize{}- Greatly limit the range of the human body's motion}\tabularnewline
				\hline 
				{\scriptsize{}Acoustic \cite{related_mc_7}} 
			  & {\scriptsize{}- Ultrasonic transmitters and receivers} {\scriptsize \par}
			  
			    {\scriptsize{}- Data-processing device}{\scriptsize \par}
			    
			  & {\scriptsize{}- Low cost}{\scriptsize \par}
				
				{\scriptsize{}- Perfectly {solves} occlusion problems} 
				
			  & {\scriptsize{}- Poor real-time capability}{\scriptsize \par}
				
				{\scriptsize{}- Easily affected by noise}\tabularnewline
				\hline 
				
			{\scriptsize{}Markerless \cite{SO_1}} 
				& {\scriptsize{}- Video and depth sensors} {\scriptsize \par}
				
				{\scriptsize{}- Data-processing device}{\scriptsize \par}
				
				& {\scriptsize{}- Low cost}{\scriptsize \par}
				
				{\scriptsize{}- Does not need wearable gears} 
				
				& {\scriptsize{}- Low accuracy}{\scriptsize \par}
				
				{\scriptsize{}- Limitation of measurement area}\tabularnewline
				\hline 	
			\end{tabular}
			\par\end{centering}
	\end{table}

2) Inertial systems apply accelerometers and gyroscopes to obtain the speed and the rotation angle of one specific point in 3D space. With the two parameters, the point displacements can be calculated during a specific period. \cite{related_mc_2}.

3) A magnetic system consists of an electromagnetic emission source and electromagnetic receivers. The magnetic source builds a low-frequency magnetic field while the user is moving in the measurement area and wearing several electromagnetic receivers. Based on the Hall effect \cite{related_mc_4}, the movement of the sensor can be estimated. For establishing a stable magnetic field while the user is moving, the electromagnetic emission source is usually mounted at the top of the measurement area.

4) Mechanical systems capture an object's movement by measuring the object's orientation and displacement with electromechanical potentiometers embedded in a mechanical exoskeleton \cite{related_mc_6}.

5) Acoustic systems consist of ultrasonic transmitters and ultrasonic receivers. By measuring the {travel time} of an ultrasonic pulse sent from the transmitters to the receivers, the distance can be estimated using the ultrasonic pulse's speed. The position then can be calculated by triangulation with multiple receivers' distance measurements from the same transmitter. By constantly calculating the positions of transmitters attached on the human body, the user's motion is captured \cite{related_mc_7}.

6) Markerless systems track a human body's motion {completely} based on {video-processing} technologies without requiring specific markers to be attached on the human body, which will be discussed in the following subsection. The comparisons of the five motion capture technologies are shown in Table \ref{tab:mc_comparisions}.

\subsection{{Markerless Systems}}
In the previous studies, the overall structure of a markerless motion capture system is generally described as having four parts: 1) {Initialization: Procedure (including calibration, model establishment, etc.) that guarantees that the system operates correctly when it is launched}; 2) {Tracking: Procedure of segmenting the human body from the background and further tracking the body in one or more frames}; 3) {Pose estimation: Procedure that identifies how the human body (individual limbs or joints) is configured in the current scene;} 4) { Pose recognition: Procedure that classifies the captured motion as one of several types of actions (such as walking, running, etc.) according to the result obtained in previous procedures \cite{intro_mc_7,intro_mc_9}.} As markerless systems realize motion capture based on the human vision principle, which mainly focuses on processing the captured color/intensity frames, {markerless systems do not require wearable devices. Hence, markerless systems are usually cost-effective and easy to set up.}

However, the methods applied in markerless systems with video-based cameras are greatly limited in terms of the following aspects: 1) human motions (or interactions) span in higher-dimensional space, while video cameras capture 2D frames, which complicates motion capture; 2) as one identical action might be performed differently at different times by one person, there exists a large number of variations of human motion, which require a robust database and {intensive match algorithm}; and 3) the methods based on one or more video cameras are sensitive to human appearance (hair, cloth, skin, etc.) and shape and lighting conditions, which makes automatic motion capture even more challenging \cite{RL_3,RL_4,RL_5}.

Different from traditional video-based cameras that capture color/intensity frames, a depth camera generates a depth map of the current scene. Using range information generated with IR light beams, depth cameras are able to work {without light requirement} and provide 3D information { without color (or texture) effects}, which largely alleviates the first and third limitations mentioned above in terms of motion capture \cite{RL_4,RL_6}. With advances in depth sensor development, depth camera prices reached consumer grade recently and are gaining popularity with time. One of the most popular motion capture systems with depth sensing features is the gait tracking system embedded in Kinect for the Windows v2 sensor \cite{001KinectWeb} by Microsoft. There are two steps for capturing human motions for gait tracking with a Kinect v2: 1) recognizing the body part from a depth frame and 2) finding the {key part's (e.g., joints)} position within one specific body part. To infer the body part, a database is built consisting of over one million depth images from a known skeleton from another motion capture system. For each real image, the body parts are rendered using computer graphic techniques. A randomized decision forest is built using the training data, and hence, the body part can be recognized after a new depth frame with unknown body parts is obtained (test data). With the recognized body parts, a mean shift algorithm is applied for 3D joint proposals. With a start-depth pixel, the mean shift algorithm helps find the modes in this density frame efficiently and finally estimates the sum of the pixel weights reaching each mode \cite{011Parts}. In this paper, based on the depth sensing technology of the Kinect sensor, we develop a three-Kinect gait tracking system with improved accuracy.

\subsection{Kinect-Based Gait Analysis}
The launch of Kinect sensors makes it possible to develop a cost-effective motion capture system \cite{RL_8}. Previous efforts focused on accessing the capability of the Kinect sensor for gait tracking in clinical applications. Both the temporal and spatial characteristics of the Kinect sensor were evaluated when motion tracking was performed on the participants who conducted specific assigned tasks. For example, to measure the Kinect sensor's temporal characteristics, a Parkinson's disease patient was asked to walk or stand on foam with one eye open while a Kinect sensor and a Vicon system tracked the patient's gait parameters simultaneously. Then, the temporal performance of the Kinect sensor is illustrated by computing the Pearson's coefficient and ICC, among others, between the tracking result of the Kinect sensor and the Vicon system, which can be considered as the standard result. 

In \cite{RL_gait_2}, the movement symptoms of the Parkinson's disease patient were assessed using the Kinect v1 sensor (i.e.,  Xbox 360 Kinect), which {achieved} good accuracy on temporal characteristics while performing poorly on spatial characteristics. Xu et al. and Auvinet et al. examined the accuracy level of the Kinect v1 sensor for estimating various gait parameters during treadmill walking with different speeds. The results indicated that the Kinect sensor has higher temporal accuracy for timing {heel} strike (HS) than for timing toe off (TO) \cite{RL_gait_3,RL_gait_4}. In \cite{RL_gait_5}, the Kinect v1 sensor was used as a clinical assessment tool for the total body center of mass (TBCM) sway measurement. The participants were asked to perform four different tasks: eyes open, eyes closed, eyes open standing on foam and eyes closed standing on foam. Compared with the Vicon system, the Kinect was more accurate in the medial-lateral direction and had better accuracy than the force plate in more challenging balance tasks.

Recently, the new Kinect v2 sensor (i.e., Xbox One Kinect sensor) was also used and assessed in gait analysis. In \cite{RL_gait_6}, the reliability and concurrent validity were evaluated while the participants were performing comfortable and fast-paced walking trials. The results showed that the Kinect v2 sensor consistently exhibited excellent concurrent validity; however, it {achieved} modest validity for medial-lateral pelvis sways. In addition, Clark et al. evaluated the reliability and concurrent validity of the Kinect v2 sensor for assessing balance control. In their experiments, the participants were assigned two tasks: static standing balance with single or double limb support conditions and dynamic balance with forward or lateral reach. The results showed that the new version of the Kinect sensor {achieves} an acceptable performance for estimating the standing balance and postural control \cite{RL_gait_1}.

In summary, previous works have demonstrated that the newly released Kinect v2 is a reliable and valid tool for gait analysis. In this paper, different from the previous study \cite{SO_1}, which improves the Kinect sensor's accuracy based on the Least Square principle and requires at least four Kinect sensors in a 3D space, we propose a {geometric} trilateration method requiring only three Kinect sensors. Based on the proposed trilateration method, an integrated gait tracking system is developed, including a new synchronization and time scheduling mechanism, which is utilized for synchronizing the data transmitted from different Kinect sensors. Furthermore, the estimated gait parameters are visualized and demonstrated based on a musculoskeletal model.

\section{System Overview}

\begin{figure}[!t]
	\begin{centering}
		\includegraphics[width=8.5cm]{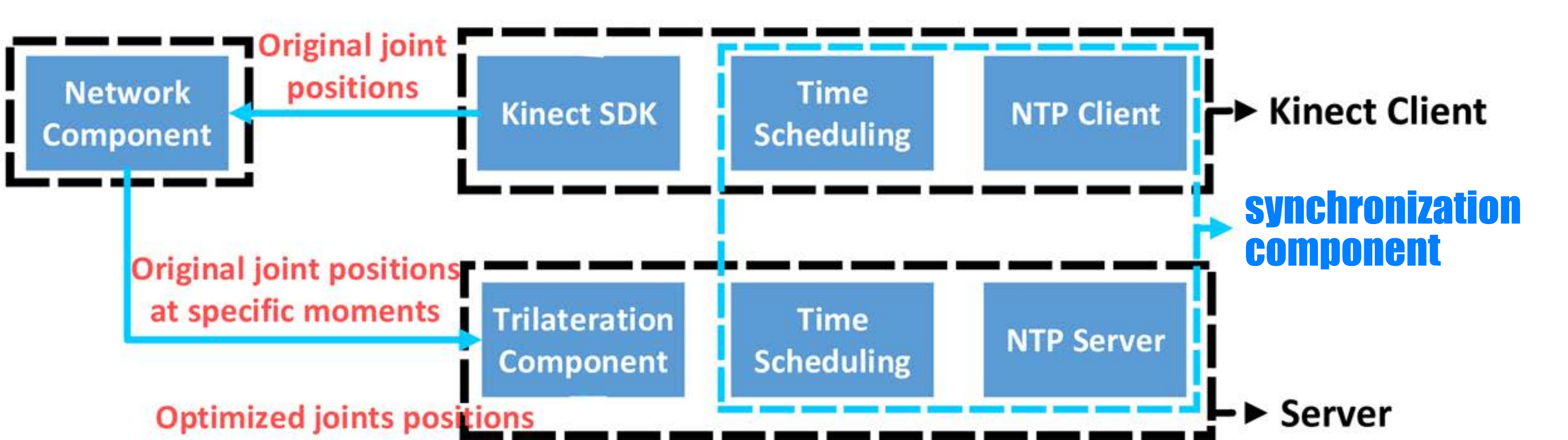}
		\par\end{centering}
	
	\protect\caption{\label{fig:architecture}Overall system architecture of a multi-Kinect human gait tracking system}
\end{figure}

Fig. \ref{fig:architecture} depicts the overall architecture of our human gait tracking system. The points tracked by the system are actually the human joint positions specified by the Kinect sensor. A total of six joint positions, including the left hip, right hip, left knee, right knee, left ankle and right ankle, are tracked. Three Kinect sensors are connected to three computers, which are {set up} as Kinect clients. Thus, three Kinect clients are built for capturing joint positions while the user is walking within a specific measurement area. This system consists of several components: 1) a synchronization component guarantees that the clocks of the Kinect clients and the server are synchronized in milliseconds by network time protocol (NTP) \cite{02time_sync}. Furthermore, it schedules the joint position transmission time sequences such that during one specific period, the server obtains the joint data one by one from each Kinect client. 2) At the Kinect client, the networking component serializes the joint data structure specified by Kinect SDK and forwards the data to the server; at the server, the network component deserializes the joint data from the three Kinect clients and sends the data to the trilateration component. 3) The trilateration component applies the trilateration method to compute the joint positions in 3D space. 

\section{Trilateration}
The basic idea of the trilateration method is utilizing multiple sensors to measure the distances between the target and each sensor and building several spheres by taking each sensor as a center point and its corresponding distance as the radius. Thus, the target position is the common interaction point of these spheres, which can be determined by solving a set of equations representing all the spheres \cite{024determination}. In the previous study, multiple Kinect sensors are applied for improving the accuracy based on {Least} Square theory, which requires additional Kinect sensors to obtain redundant data for minimizing the error \cite{SO_1}.  {The Least square method is one of the implementations based on the trilateration principles, and it applies difference measurements (leading to the one more Kinect sensors requirement for 3D scenarios) in the set of least square differential equations. In this paper, by applying geometric principles, we simplify the trilateration computation by calculating the measurement data directly from 3 Kinects (one less Kinect compared with the case of least square trilateration) to obtain the same accuracy and also greatly decreasing the complexity of the computation/system.} The geometry-based trilateration method (locating the target based on geometric and mathematical rules) is applied with three Kinect v2 sensors for tracking human gait in 3D. An experiment is conducted for verifying that the mentioned trilateration method improves the Kinect sensor's accuracy.

\subsection{\label{sc:Multi-KinectTrilateration}Multi-Kinect Trilateration in 3D Space}

\begin{figure*}[!t]
	\begin{centering}
		\includegraphics[height=9cm]{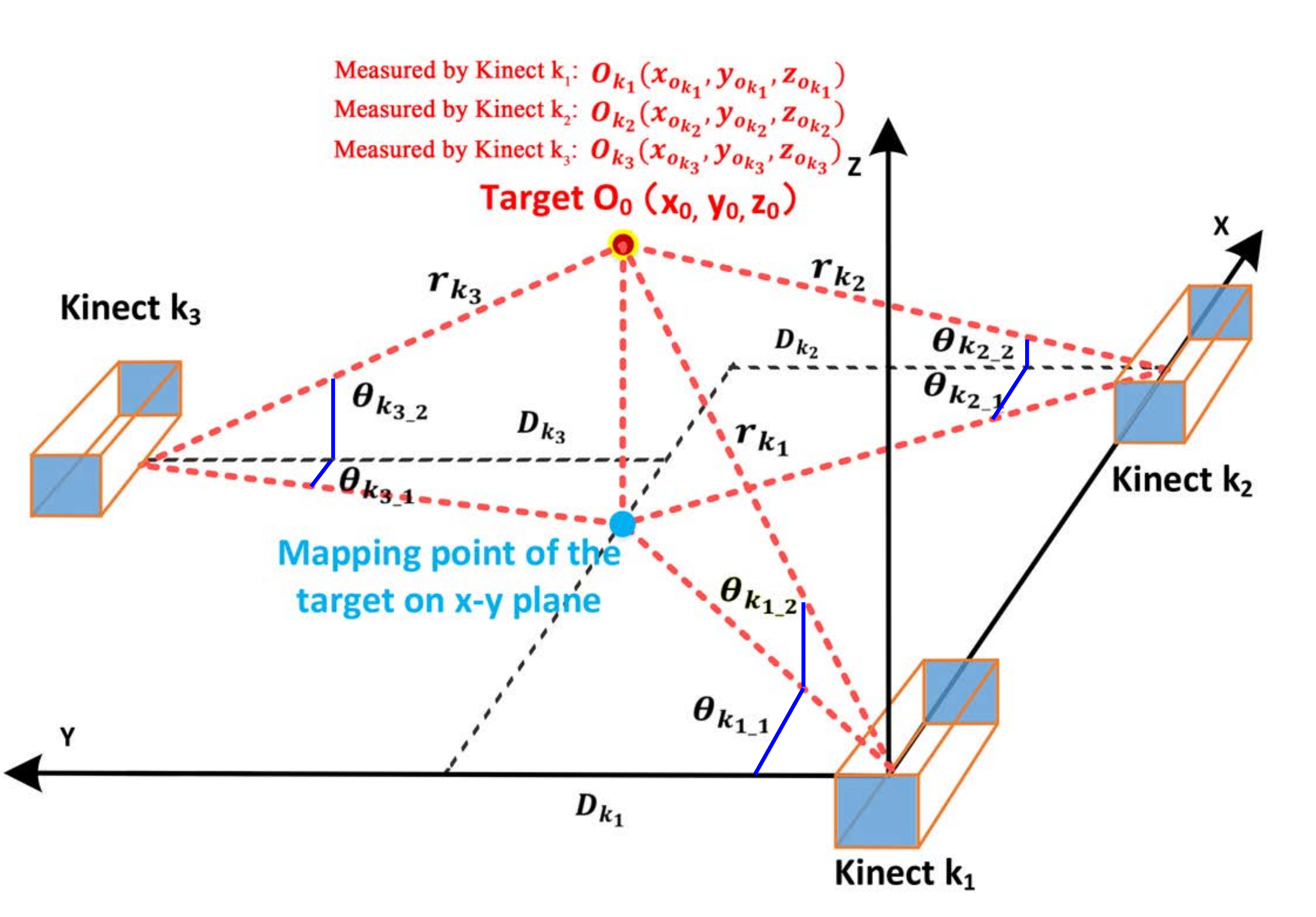}
		\par
	\protect\caption{\label{fig:multi-Kinect}{Trilateration set-up of three Kinects within a coordinate system.}}
\end{centering}
\end{figure*} 

We apply the trilateration principle {with} a multiple-Kinect set up to improve the location accuracy
of a target object when it is placed where the Kinect sensors cannot generate depth values accurately (e.g., distances larger than 4 m from the observing sensor). The three-Kinect set up is shown in \figurename \ref{fig:multi-Kinect}. In this configuration, the target object is the central area of a planar surface located more than 4 m away from the three Kinect sensors\cite{SO_1}. The three vertexes $k_{1}(x_{k_{1}},y_{k_{1}},z_{k_{1}})$, $k_{2}(x_{k_{2}},y_{k_{2}},z_{k_{2}})$ and $k_{3}(x_{k_{3}},y_{k_{3}},z_{k_{3}})$ representing the locations of the three Kinects are positioned at the same horizontal plane ($z_{k_1}=z_{k_2}=z_{k_3}= 0$) and $O_{0}(x_{0},y_{0},z_{0})$ representing the location of the target object positioned above the three Kinect sensors ($z_{0}>0$). $l_{k_{1}k_{2}}$, $l_{k_{1}k_{3}}$, $l_{k_{2}k_{3}}$ are real-world measurements, where $l_{k_{i}k_{j}}=||k_{i}-k_{j}||_2$ ($i,j=1,2,3\; and\; i\neq j$) is the distance between Kinect $k_{i}$ and Kinect $k_{j}$. The depth value and two angles $(\:D_{k_{1}}, \:\theta_{k_{1\_1}}, \:\theta_{k_{1\_2}})$, ($\:D_{k_{2}}, \:\theta_{k_{2\_1}}, \:\theta_{k_{2\_2}})$ and $(\:D_{k_{3}}, \:\theta_{k_{3\_1}}, \:\theta_{k_{3\_2}})$, as shown in Fig. \ref{fig:multi-Kinect}, are measured by Kinects $k_{1}$, $k_{2}$ and $k_{3}$, respectively. {The depth values $D_{k_{1}}$,  $D_{k_{2}}$ and $D_{k_{3}}$, which denote the distance from the target object to the vertical plane of the Kinect, are calculated by averaging 20$\times$20 pixel values in the depth image at the position of the target object.}

The input of the trilateration includes the measured parameters ($l_{k_{1}k_{2}}$, $l_{k_{1}k_{3}}$ and $l_{k_{2}k_{3}}$) and the parameters obtained by each Kinect sensor ($\:D_{k_{1}}, \:\theta_{k_{1\_1}}, \:\theta_{k_{1\_2}}$), ($\:D_{k_{2}}, \:\theta_{k_{2\_1}}, \:\theta_{k_{2\_2}}$) and ($\:D_{k_{3}}, \:\theta_{k_{3\_1}}, \:\theta_{k_{3\_2}}$), while the output is the optimized position of $O_0$ (denoted as $O^{'}$). For generating the spheres' equations and further calculating the optimized position of $O$, the vertexes' positions and the displacements of each Kinect sensor and the target are required.

To achieve the vertexes' positions, the coordinates of the three Kinect sensors are calculated using Equation \ref{eq:vertexes}.
\begin{equation}
\label{eq:vertexes}
\begin{array}{l}
k_{1}\left\{
\begin{array}{l}
x_{k_{1}}=0\\y_{k_{1}}=0\\z_{k_{1}}=0\\
\end{array}\right.\\
k_{2}\left\{
\begin{array}{l}
x_{k_{2}}=l_{k_{1}k_{2}}\\y_{k_{2}}=0\\z_{k_{2}}=0\\
\end{array}\right.\\
k_{3}\left\{
\begin{array}{l}
x_{k_{3}}=\sqrt{l_{k_{1}k_{3}}^{2}-y_{k_{3}}^{2}}\\
y_{k_{3}}=\frac{2\sqrt{S_{1}(S_{1}-l_{k_{1}k_{2}})(S_{1}-l_{k_{1}k_{3}})(S_{1}-l_{k_{2}k_{3}})}}{l_{k_{1}k_{2}}}\\
z_{k_{3}}=0\\
\end{array}\right.\\
\end{array}
\end{equation}

Similarly, based on the geometry, the displacements (the radii for building the spheres' equations) between each Kinect sensor and the target object are calculated using Equation \ref{eq:radiuses}.
\begin{equation}
\label{eq:radiuses}
 \left\{
\begin{array}{l}
r_{k_{1}}=\sqrt{(\frac{D_{k_{1}}}{\text{cos}\:\theta_{k_{1\_1}}})^{2}+(D_{k_{1}}\cdot \text{tan}\:\theta_{k_{1\_2}})^{2}}\\
r_{k_{2}}=\sqrt{(\frac{D_{k_{2}}}{\text{cos}\:\theta_{k_{2\_1}}})^{2}+(D_{k_{2}}\cdot \text{tan}\:\theta_{k_{2\_2}})^{2}}\\
r_{k_{3}}=\sqrt{(\frac{D_{k_{3}}}{\text{cos}\:\theta_{k_{3\_1}}})^{2}+(D_{k_{3}}\cdot \text{tan}\:\theta_{k_{3\_2}})^{2}}\\
\end{array}
\right.
\end{equation}

Using the results from Equation \ref{eq:vertexes} and {Equation} \ref{eq:radiuses}, the three spheres' equations (the three Kinect sensors' positions as the centers and displacements as radii, respectively) are built as follows
\begin{equation}
\label{eq:spheres}
 \left\{
\begin{array}{c}
(x_{O^{'}}-x_{k_{1}})^{2}+(y_{O^{'}}-y_{k_{1}})^{2}+(z_{O^{'}}-z_{k_{1}})^{2}=r_{k_{1}}^{2} \\
(x_{O^{'}}-x_{k_{2}})^{2}+(y_{O^{'}}-y_{k_{2}})^{2}+(z_{O^{'}}-z_{k_{2}})^{2}=r_{k_{2}}^{2} \\
(x_{O^{'}}-x_{k_{3}})^{2}+(y_{O^{'}}-y_{k_{3}})^{2}+(z_{O^{'}}-z_{k_{3}})^{2}=r_{k_{3}}^{2} \\
\end{array}
\right.
\end{equation}
where $z_{k_{1}}=z_{k_{2}}=z_{k_{3}}=0$, as Kinects v2 $k_{1}$, $k_{2}$ and $k_{3}$ are set to the same horizontal plane. Thus, the target's trilateration result ($O^{'}(x_{O^{'}}$, $y_{O^{'}}$ and $z_{O^{'}})$) can be calculated by solving Equation \ref{eq:spheres}.
The results are shown below in Equation \ref{eq:result}.
{
	\begin{equation}
		\label{eq:result}
		\left\{
		\begin{array}{l}
			x_{O^{'}}=\frac{r_{k_{1}}^{2}-r_{k_{2}}^2+x_{k_{2}}^{2}}{2x_{k_{2}}} \\
			y_{O^{'}}=\frac{r_{k_{1}}^{2}-r_{k_{3}}^2+x_{k_{2}}^{2}+y_{k_{3}}^{2}-x_{k_{3}}(\frac{r_{k_{1}}^{2}-r_{k_{2}}^2+x_{k_{2}}^{2}}{x_{k_{2}}})}{2y_{k_{3}}} \\
			z_{O^{'}}=\pm\sqrt{r_{k_{1}}^{2}-x_{O^{'}}^{2}-y_{O^{'}}^{2}} \\
		\end{array}
		\right.
	\end{equation}
	The result shows two solutions of z parameter. To determine the actual solution, all Kinects are fixed in a horizontal plane of which all corresponding joints are in the same side. By way of illustration, the minus solution would be chosen if all joints are above the plane.
}
\subsection{\label{sc:Multi-Kinect Trilateration in Human Gait Tracking System}Multi-Kinect Trilateration in the Human Gait Tracking System}
\begin{figure}[!t]
\begin{centering}
		\includegraphics[width=9cm]{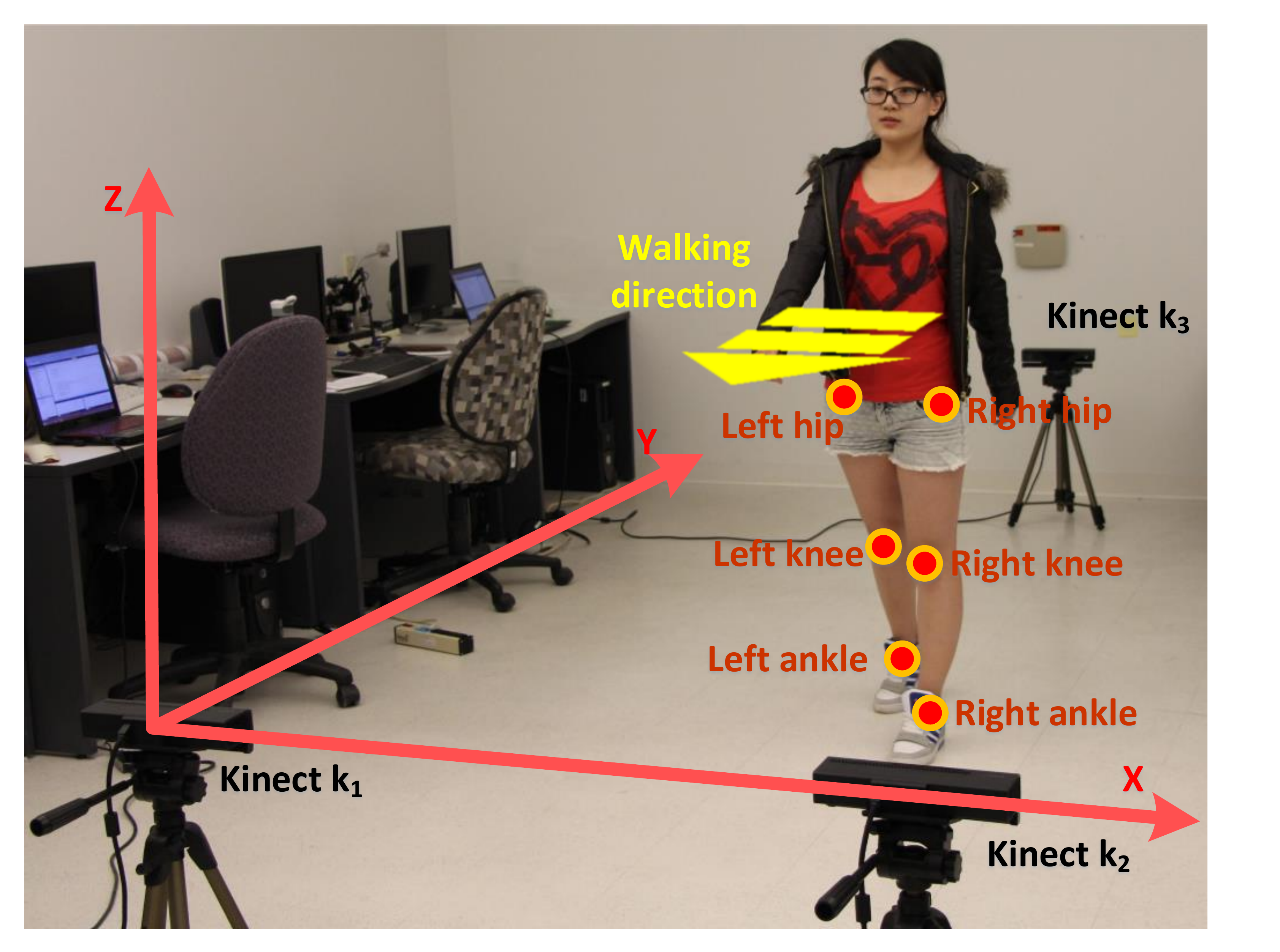}
		\par
	\protect\caption{\label{fig:trilaterationsetup} Trilateration configuration in the experiment: the corresponding joints, coordinate system, walking direction and positions of the Kinects are shown.}
\end{centering}
\end{figure}

This system {applies} the multi-Kinect trilateration principle mentioned in Section \ref{sc:Multi-KinectTrilateration} to locate joint positions in space. The participant is required to walk for several steps in a specific direction inside the triangle with three Kinect sensors as the vertexes.  A total of six joints including the right hip, left hip, right knee, left knee, right ankle and left ankle are tracked while the user is walking. The coordinate setting is shown in \figurename \ref{fig:trilaterationsetup}.

According to the manual, the Kinect v2 sensor cannot observe anything when the target is positioned less than 0.5 m away from the sensor. Hence, the entire walking range is designed as a 4.6-m-height triangle, and the recommend walking area is a 3-m straight line at the central part. As the Kinect sensor can hardly track the human body when the distance is larger than 10 m, the recommended configuration for trilateration is that the largest distances between each set of two Kinect sensors should be less than 10 m. The trilateration method would be executed once three different joint data structures are received by the server in the right order, i.e., the trilateration thread would be executed after the other three threads obtain three joint data structures from the three Kinect sensors {properly}. The total time cost for getting one optimized joint position depends on the environmental conditions. Here, under our configuration, the time cost for one optimized joint position is 60 ms. The system calculates 1,000 optimized positions for one specific joint (6,000 optimized joint positions for six joints) while the participant is walking within the 3-m range. Thus, the total duration that the user needs to walk for is 60 s. To walk within a 3-m range for 60 s, the user might need to walk 3 m and walk back to the origin several times. The duration can also be customized manually if more joint data are required (e.g., a treadmill is applied to obtain more gait parameters for further analysis).

\section{Synchronization}
Once the joint positions are obtained by the three Kinect sensors simultaneously, the joint positions need to be forwarded to the server for optimization with trilateration calculations. Various factors may affect the time when the server receives the joint data. These factors include the hardware differences (even the same specifications), the network conditions, and lighting conditions, among others. The speeds of receiving joint data from the Kinect clients are different (i.e., during one specific period, joint data from one or two Kinect clients may be blocked on the network), which can lead to an incorrect trilateration result. To tackle this problem, a time-scheduling component is developed to control the order in which the data of the Kinect clients are sent. Furthermore, to guarantee the time-scheduling component's correctness, the clocks of the computers (the server and three Kinect clients) are synchronized with a synchronization component based on NTP. As the general time cost for one Kinect sensor to generate joint data and the server to receive it is 3 to 5 ms (measured with Kinect SDK and a C++ Server), the synchronization component must have a high resolution and accuracy in milliseconds.

{Because Kinect uses USB 2.0 as the interface to connect with the computer, the latency of transmitting data from the Kinect to the computer has to be addressed.  According to the official technical specification{\cite{usb_1}}, the interval of USB 2.0 data transmission is 0.125 ms, indicating that the time delay from the Kinect to computer is less than 0.125 ms. In our system, the time interval for obtaining data from the Kinect client is 15 ms, which is much larger than 0.125 ms. Thus, the time delay caused by data transmission by USB 2.0 would not affect the orders of frames from different Kinect sensors.}

\subsection{Clock Synchronization}
{Currently, PTP{\cite{ieee_1588_1}} and NTP \cite{02time_sync} are the most widely used synchronization protocols in the computer network field. There are many implementations of these protocols such as Chrony, NetTime, and Atomic Clock Sync, among others. The reason that we chose to build a customized NTP synchronization scheme in our system rather than using off-the-shelf implementations is reducing both the complexity of the system and the usage of resources: 1) most implementations such as Chrony are designed to work in a network with multiple servers in a complex structure. The network structure of our system is relatively simple, allowing us to build a customized NTP implementation that more precisely fits the requirement. 2) In contrast, compared with NTP, PTP is designed to achieve clock accuracy at the sub-microsecond level, which requires more network resources. However, the accuracy requirement for time synchronization in our proposed system is at the millisecond level (1 ms), which is 100 times lower than the accuracy provided by PTP (0.01 ms).}

To synchronize clocks of different devices based on NTP, the NTP client (the target to be synchronized) sends a NTP message to the NTP server (the reliable time source) attached with a timestamp $T_{1}$ when the message leaves the NTP client. Second, the NTP server receives the NTP message and attaches a timestamp ($T_{2}$) when the message arrives at the NTP server. Third, the NTP server sends back the NTP message to the NTP client attached with a timestamp $T_{3}$ when the message leaves the NTP server. Finally, the NTP client receives the message attached with a timestamp $T_{4}$ when the message arrives at the NTP client. With the four timestamps, we have
\begin{equation}
\label{eq:ntp1}
 \left\{
\begin{array}{l}
T_{2}=T_{1}+t+d_{1}\\
T_{4}=T_{3}-t+d_{2}\\ 
d=d_{1}+d_{2}\\
\end{array}
\right.
\end{equation}
where $t$ is the time offset between the NTP client's clock and the NTP server's clock, $d_{1}$ is the transmission delay while the message is being sent from the NTP client to the NTP server, $d_{2}$ is the transmission delay while the message is being sent back to the NTP client from the NTP server, and $d$ is the total transmission delay of the round trip. By assuming $d_{1}=d_{2}$, we have
\begin{equation}
\label{eq:ntp2}
 \left\{
\begin{array}{l}
t=\frac{(T_{2}-T_{1})-(T_{4}-T_{3})}{2}\\
d=(T_{2}-T_{1})+(T_{4}-T_{3})\\ 
\end{array}
\right.
\end{equation}
According to Equation \ref{eq:ntp1} and Equation \ref{eq:ntp2}, we have
\begin{equation}
\label{eq:ntp3}
t=(T_{2}-T_{1})+d_{1}=(T_{2}-T_{1})+\frac{d}{2}
\end{equation}
We have the following: 1) $t$ or $d$ is relevant to the differences of $T_{2}-T_{1}$ and $T_{4}-T_{3}$, which means that $t$ and $d$ are not relevant to the process delay ($T_{3}-T_{2}$) of the NTP server; 2) Equation \ref{eq:ntp2} is obtained when $d_{1}=d_{2}$ ({in fact, $d_{1}$ and $d_{2}$ vary in the range of $[0, d]$}). According to Equation \ref{eq:ntp3}, the maximum error of $t$ is $\pm\frac{d}{2}$.

Usually, in a local area network (LAN), the transmission delay ($d$) is less than 1 ms (in our proposed system, the transmission delay is less than 1 ms as tested by the {synchronization component}). Thus, the synchronization error of the synchronization component would theoretically be less than 0.5 ms (less than $\pm\frac{d}{2}$), which satisfies the requirement of the time scheduling component.

\subsection{\label{sc:Time Scheduling}Time Scheduling}
{As clients send joint data frames continuously, the server receives joint data one by one,} and there is always a data receiving sequence for the three Kinect clients. Due to the limitations of the environmental conditions and hardware differences, the joint data {cannot} arrive in the right order (i.e., joint data from one or two Kinect clients might be blocked on the network while one or two of the rest of the Kinect clients are sending more joint data to the server at the same time). 

\begin{figure}[!t]
	\begin{centering}
		\subfloat[\label{fig:timeschedule2}]{\begin{centering}
				\includegraphics[width=8.5cm]{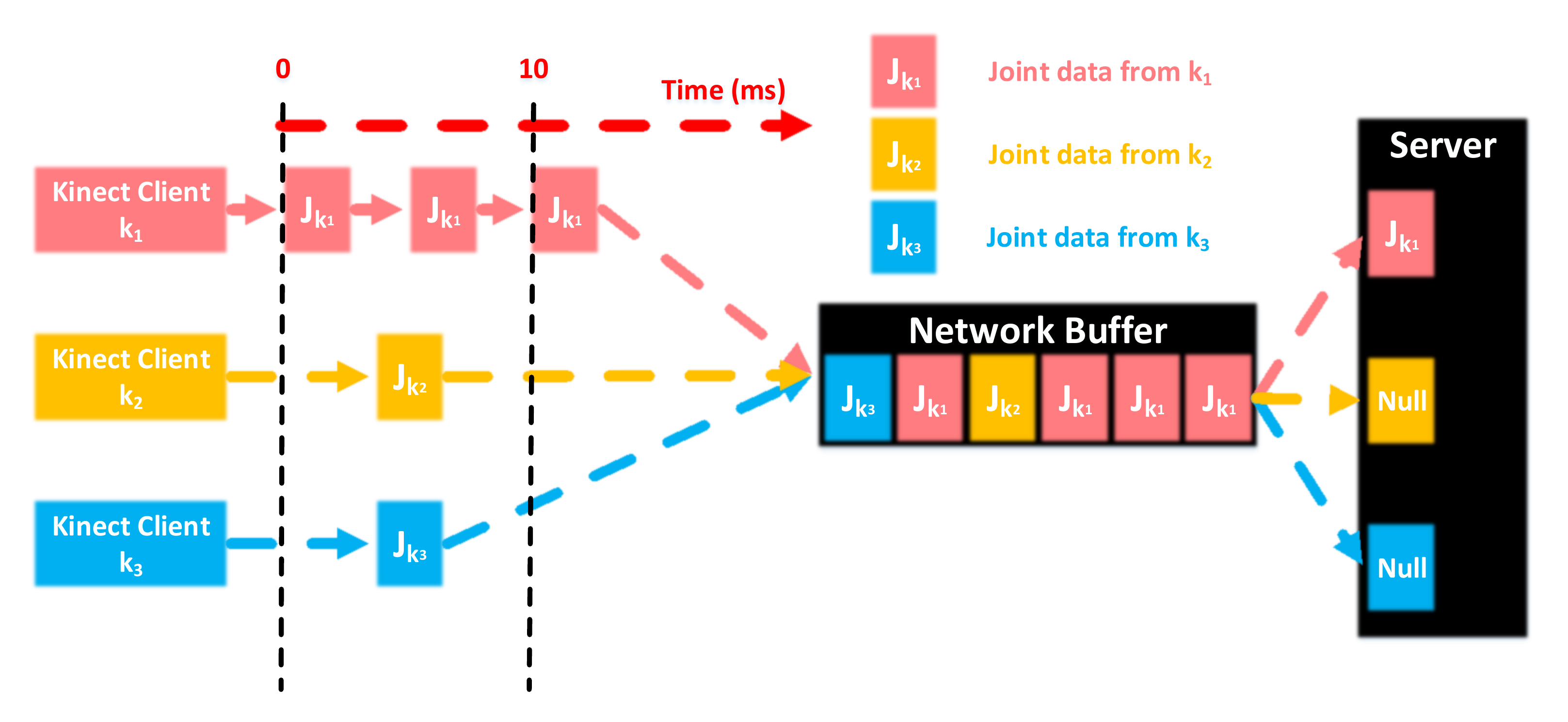}
				\par\end{centering}
		}
		\par\end{centering}
	\begin{centering}
		\subfloat[\label{fig:timeschedule}]{\begin{centering}
				\includegraphics[width=8.5cm]{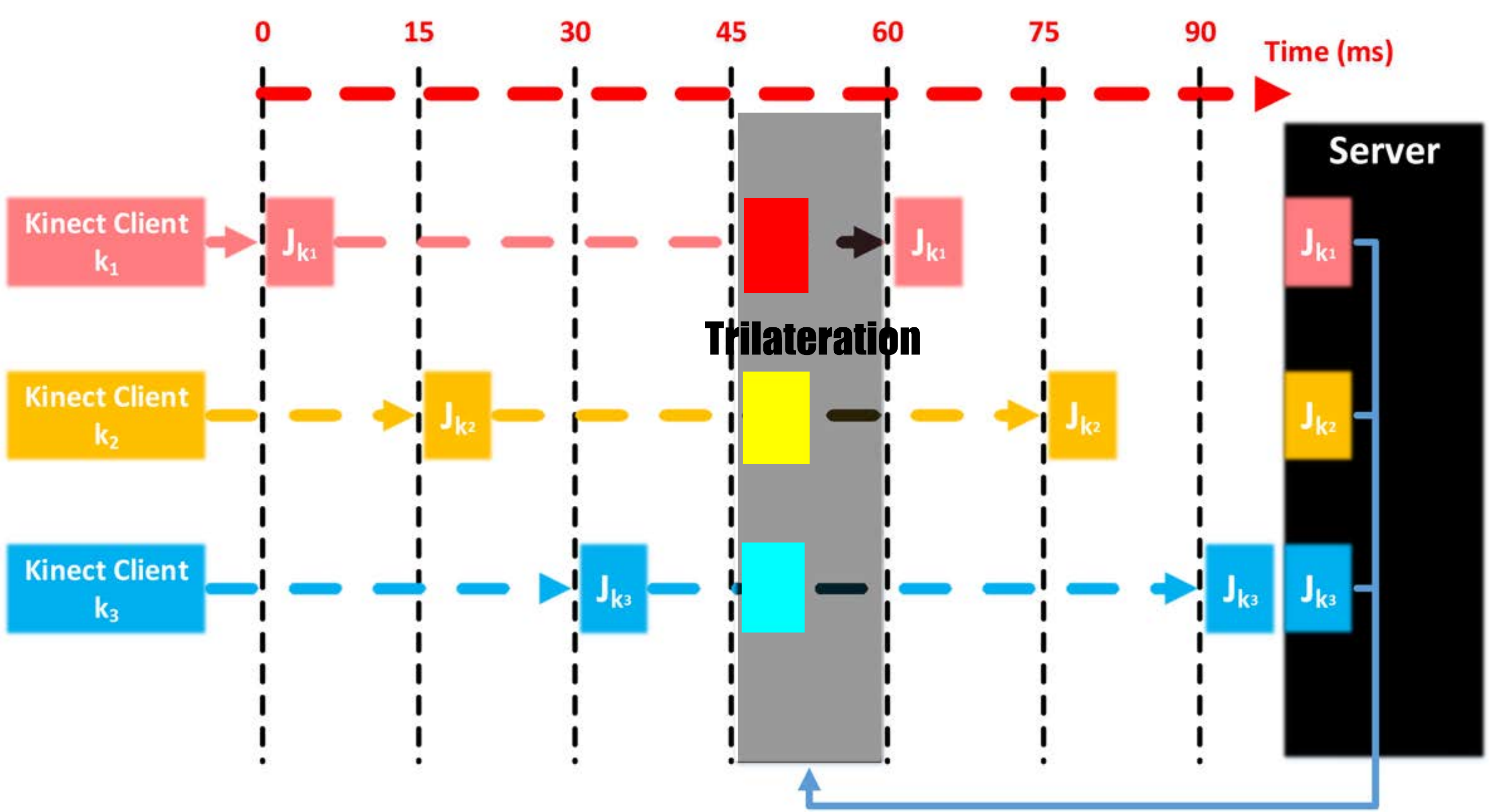}
				\par\end{centering}
		}
		\par\end{centering}
	\protect\caption{Time schedule. (a) Joint data from Kinect client $k_{1}$ and Kinect client $k_{2}$ are blocked on the network. (b) To tackle the problem corresponding to (a), a time schedule is designed for timing the Kinect clients' data transmission.}
\end{figure}

{In} Fig. \ref{fig:timeschedule2}, the speed of receiving joint data from Kinect client $k_{1}$ is much faster than that from Kinect client $k_{2}$ and Kinect client $k_{3}$. The joint data from $k_{2}$ and $k_{3}$ are blocked on the network buffer, and the server does not receive the joint data from $k_{2}$ and $k_{3}$ within a short time, which may lead to an incorrect trilateration result or a lack of enough data for the trilateration procedure.

Instead of continuing to send joint data to the server from Kinect clients, the time-scheduling component controls the Kinect clients' data-sending sequence by making Kinect clients obtain and send joint data one by one. Though the time cost of generating joint data by one specific Kinect client and receiving the data by the server is 3 to 5 ms, the time cost of trilateration is less than 5 ms. {Throughout the experiment, we found that this component cannot realize 5-ms resolution time scheduling, as the clock resolution of Windows API is unstable (e.g., if Kinect client $k_{1}$ obtains joint data at time $T_{1}$ and sends it to the server, the time-scheduling component cannot guarantee that Kinect client $k_{2}$ would obtain and send the joint data to the server at $T_{1}$+5 ms, as the time resolution is in the range of 1 to 15 ms).} This is caused by the overclocking feature {of} most current CPUs \cite{gaitsys_ts_1}. To guarantee that the Kinect clients obtain joint data one by one, this time schedule component schedules the periods (15-ms periods) for different Kinect clients to send joint data, as shown in Fig. \ref{fig:timeschedule}: the time line is divided into 15-ms periods, and each 15-ms period is designed for a specific Kinect client to send joint data or for trilateration calculations to optimize the joint position. Hence, each set of four 15-ms periods form a 60-ms iteration. During the 60-ms iteration, three joint data sets from three different Kinect clients are received by the server. Then, an optimized joint position is calculated by the trilateration method as three joint data sets are supposed to be received by the server during the previous three 15-ms periods. { The Kinects obtain data for the next iteration at the same time as when the server conducts the trilateration procedure. }


\section{System Results}
\subsection{Verification of Multi-Kinect Trilateration}
To verify that the trilateration method improves the overall Kinect measurements, three
calculated positions of $O$ (i.e., $O_{k_{1}}(x_{O_{k_{1}}},\, y_{O_{k_{1}}}, \,z_{O_{k_{1}}})$, $O_{k_{2}}(x_{O_{k_{2}}},\, y_{O_{k_{2}}}, \,z_{O_{k_{2}}})$, $O_{k_{3}}(x_{O_{k_{3}}},\, y_{O_{k_{3}}}, \,z_{O_{k_{3}}})$) in Fig. \ref{fig:multi-Kinect} are considered
\begin{equation}
\label{eq:relocate}
\begin{array}{l}
O_{k_{1}}\left\{
\begin{array}{l}
x_{O_{k_{1}}}=D_{k_{1}}\cdot \text{tan}\:\theta_{k_{1\_1}}\\y_{O_{k_{1}}}=D_{k_{1}}\\z_{O_{k_1}}=D_{k_{1}}\cdot \text{tan}\:\theta_{k_{1\_2}}\\
\end{array}\right.\\
O_{k_{2}}\left\{
\begin{array}{l}
x_{O_{k_{2}}}=x_{k_2}-D_{k_{2}}\cdot \text{tan}\:\theta_{k_{2\_1}}\\y_{O_{k_{2}}}=D_{k_{2}}\\z_{O_{k_2}}=D_{k_{2}}\cdot \text{tan}\:\theta_{k_{2\_2}}\\
\end{array}\right.\\
O_{k_{3}}\left\{
\begin{array}{l}
x_{O_{k_{3}}}=x_{k_3}-D_{k_{3}}\cdot \text{tan}\:\theta_{k_{3\_1}}\\y_{O_{k_{3}}}=y_{k_3}-D_{k_{3}}\\z_{O_{k_3}}=D_{k_{3}}\cdot \text{tan}\:\theta_{k_{3\_2}}\\
\end{array}\right.
\end{array}
\end{equation}
where $O_{k_{1}}$, $O_{k_{2}}$ and $O_{k_{3}}$ are calculated solely based on $k_{1}$, $k_{2}$ and $k_{3}$, respectively. For example, $O_{k_{1}}(x_{O_{k_{1}}},\, y_{O_{k_{1}}},\,z_{O_{k_1}})$ is calculated with the Kinect $k_{1}$'s position $k_{1}(x_{k_{1}},\, y_{k_{1}},\,z_{k_1})$ and its corresponding depth value and two angles $(D_{k_{1}},\theta_{k_{1\_1}},\theta_{k_{1\_2}})$ using geometry. Thus, a total of four calculated positions of $O$ (i.e., $O_{k_{1}}$, $O_{k_{2}}$, $O_{k_{3}}$ and $O^{'}$) are compared.

{As for the multi-Kinect trilateration system's structure }described in Section \ref{sc:Multi-KinectTrilateration}, three Kinect v2 sensors are positioned at each vertex of a isosceles triangle. Without loss of generality, we set the location of $O$ to be in front of Kinect v2 $k_{3}$. The radii ($r_{1}$, $r_{2}$ and $r_{3}$) and the vertex positions ($k_{1}(x_{k_{1}},\:y_{k_{1}})$, $k_{2}(x_{k_{2}},\:y_{k_{2}})$, $k_{3}(x_{k_{3}},\:y_{k_{3}})$) are calculated first, and then the "trilaterate" position ($O^{'}$) is computed using the steps described in Section \ref{sc:Multi-KinectTrilateration} as shown in Table \ref{tab:Trilateration-variables}.

\begin{table}[!t]
	\protect\caption{\label{tab:Trilateration-variables} Localization Comparison of a Single Kinect and Multi-Kinect Trilateration}
	\begin{centering}
		\begin{tabular}{|>{\centering}p{1.7cm}|>{\centering}p{2.3cm}|>{\centering}p{3.5cm}|}
			\hline 
			{\scriptsize{}Method} & {\scriptsize{}Measured/Calculated Position of $O$} & {\scriptsize{}Measurement/Calculation Error (Difference between  $O_{o}$ and the Measured/Calculated Position of $O$) }\tabularnewline
			\hline 
			\hline 
			{\scriptsize{}Measurement of Kinect $k_{1}$}  & {\scriptsize{}$O_{k_{1}}$}{\scriptsize \par}
			
			{\scriptsize{}(928.7, 4042.3, 407.3)} & {\scriptsize{}$|O_{0}-O_{k_{1}}|=18.88\:$mm}\tabularnewline
			\hline 
			{\scriptsize{}Measurement of Kinect $k_{2}$}  & {\scriptsize{}$O_{k_{2}}$}{\scriptsize \par}
			
			{\scriptsize{}(901.3, 4045.4, 404.3)} & {\scriptsize{} $|O_{0}-O_{k_{2}}|=17.73\:$mm}\tabularnewline
			\hline 
			{\scriptsize{}Measurement of Kinect $k_{3}$}  & {\scriptsize{}$O_{k_{3}}$}{\scriptsize \par}
			
			{\scriptsize{}(915, 4047.6, 431.5)} & {\scriptsize{}$|O_{0}-O_{k_{3}}|=22.74\:$mm}\tabularnewline
			\hline 
			{\scriptsize{}Multi-Kinect Trilateration} & {\scriptsize{}$O^{'}$}{\scriptsize \par}
			
			{\scriptsize{}(909.0, 4045.9, 415.5)} & {\scriptsize{}$|O_{0}-O^{'}|=12.20\:$mm}\tabularnewline
			\hline 
		\end{tabular}
		
		\begin{tablenotes}
			\item[*]* The measured position of $O_{o}$ is (915, 4055, 410).
		\end{tablenotes}
		
		\par\end{centering}
\end{table}
To verify the validity of the multi-Kinect trilateration, the position of $O$ (denoted as $O_o$) is measured {using a laser distance meter (distance measurement precision: $\pm$2 mm)}. $|O_{k_{1}}-O_{o}|$, $|O_{k_{2}}-O_{o}|$ and $|O_{k_{3}}-O_{o}|$ are compared with $|O^{'}-O_{o}|$, where $O_{k_{1}}$, $O_{k_{2}}$ and $O_{k_{3}}$ are obtained solely based on the measurements from Kinects $k_{1}$ to Kinect $k_{3}$, respectively. $O^{'}$ is computed based on the measurements from the three Kinects. The four measurement errors are compared, and {the result }shows that the position of $O$ observed by the {trilateration} method has the smallest measurement error (12.20 mm); the position of $O$ observed by Kinects $k_{1}$, $k_{2}$ and $k_{3}$ {have} larger measurement errors (18.88 mm for Kinect $k_{1}$, 17.73 mm for Kinect $k_{2}$ and 22.44 mm for Kinect $k_{3}$), which indicates that the trilateration method has the best result based on the parameters obtained from the three {Kinect} sensors.

\subsection{Gait Tracking}
\begin{figure}[!t]
	\begin{centering}
		\includegraphics[width=9cm]{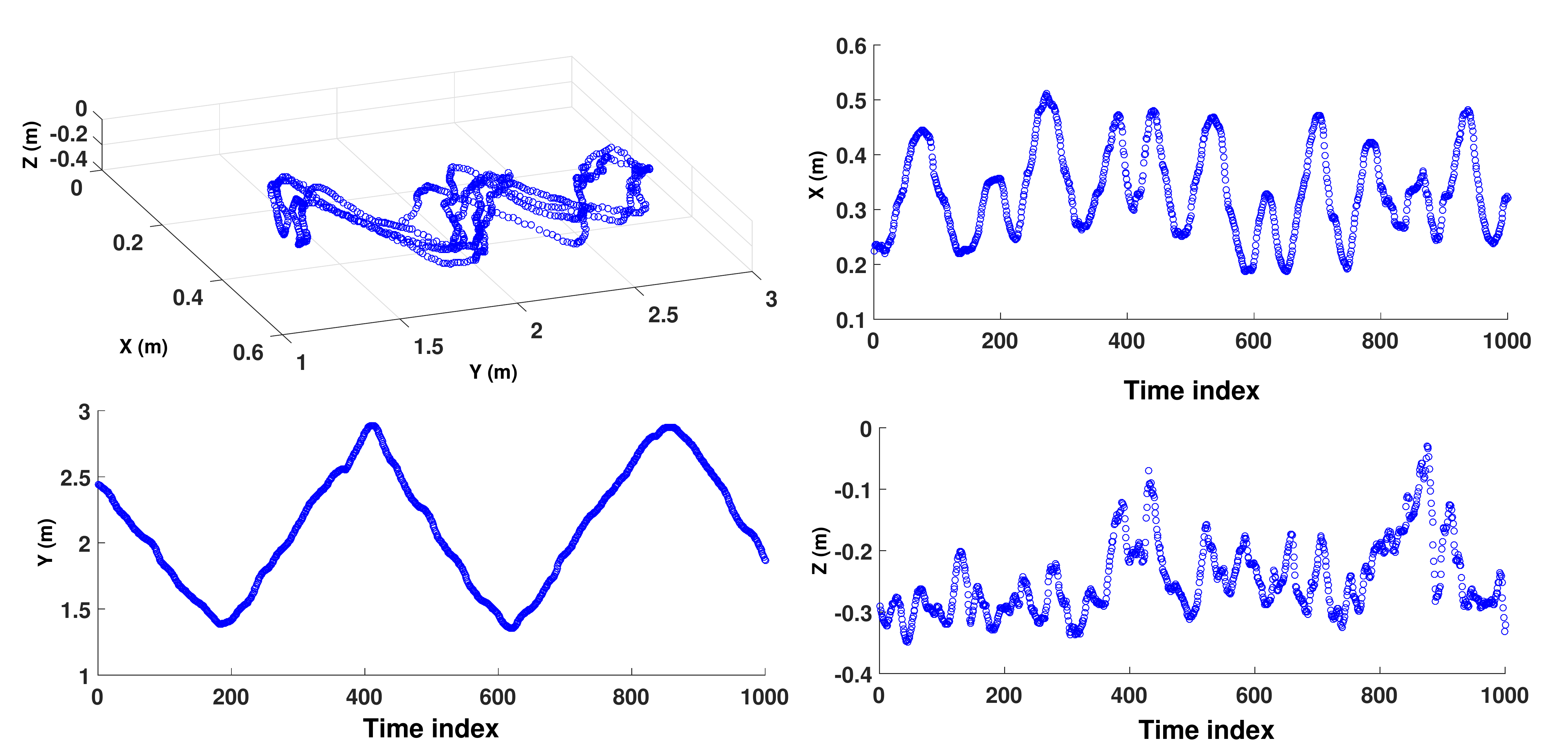}
		\par\end{centering}
	
	\protect\caption{\label{fig:gaitsys_sysres} Trilateration results for the left knee. The top-left sub-figure shows how the left knee joint is moving in the designed 3D coordinate system as shown in Fig. \ref{fig:trilaterationsetup}  with 1000 recorded position {movements} while the participant is walking. The top-right, bottom-left and bottom-right figures {show} how the joint is moving in one specific direction corresponding to the X, Y and Z axes, respectively, in Fig. \ref{fig:trilaterationsetup}.}
\end{figure}

In this human gait tracking system, to track joints with time, a total of 1000 positions are recorded during 30 s, {which contains all observed gait cycles} (i.e., 6000 positions for six joints are recorded during 30 s while the user is walking within the measurement area). In our set up (mentioned in Section \ref{sc:Multi-Kinect Trilateration in Human Gait Tracking System}), without loss of generality, the user is required to walk three steps forward, four steps backward, four steps forward, five steps backward, and three steps forward. The trilateration result of the six joint movements is then recorded in a text file. For more details on the human gait movement, the text file is loaded into MATLAB and analyzed.

As shown in \figurename \ref{fig:gaitsys_sysres}, 1,000 {discrete} positions for each knee are recorded for 30 s and then visualized using MATLAB. In this figure, the X, Y, Z axes correspond to the three directions of the trilateration configuration mentioned in \figurename \ref{fig:trilaterationsetup}. As the user walks forward and backward several times, the movements in the 3D coordinate system (the upper-left sub-figure of  \figurename \ref{fig:gaitsys_sysres}) might not be clear, while the last three sub-figures show the details on how the joint is moving in specific directions. For example, in \figurename \ref{fig:gaitsys_sysres}, the bottom-left sub-figure shows how the user's left knee is moving in the Y direction, which indicates that the user's left knee moves in the direction opposite to the Y axis for about 1.5 m during the first 200 position {movements}, 2.5 m in the direction of the Y axis during the following 200 position {movements}, and, similarly, the knee move forwards or backwards several times. {The data produced by} evaluating these joint movements can be applied for entertainment or medical purposes.

\subsection{Time Synchronization}
\begin{figure}[!t]
	\begin{centering}
		\subfloat[]{\begin{centering}\label{fig:error_kinects_1}
				\includegraphics[width=8cm]{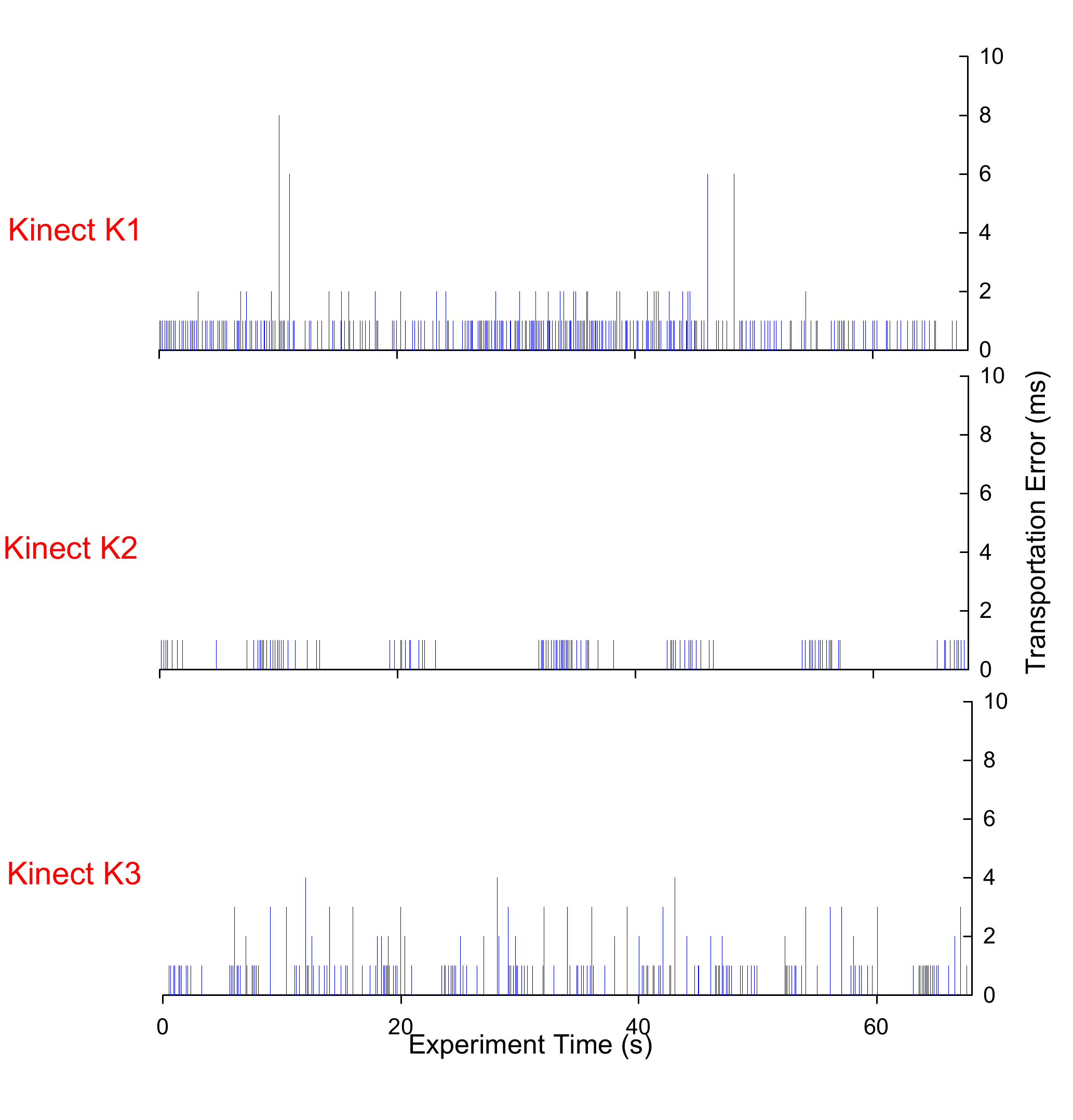}
				\par\end{centering}
		}
		\par\end{centering}
	
	\begin{centering}
		\subfloat[]{\begin{centering}\label{fig:error_kinects_2}
				\includegraphics[width=8cm]{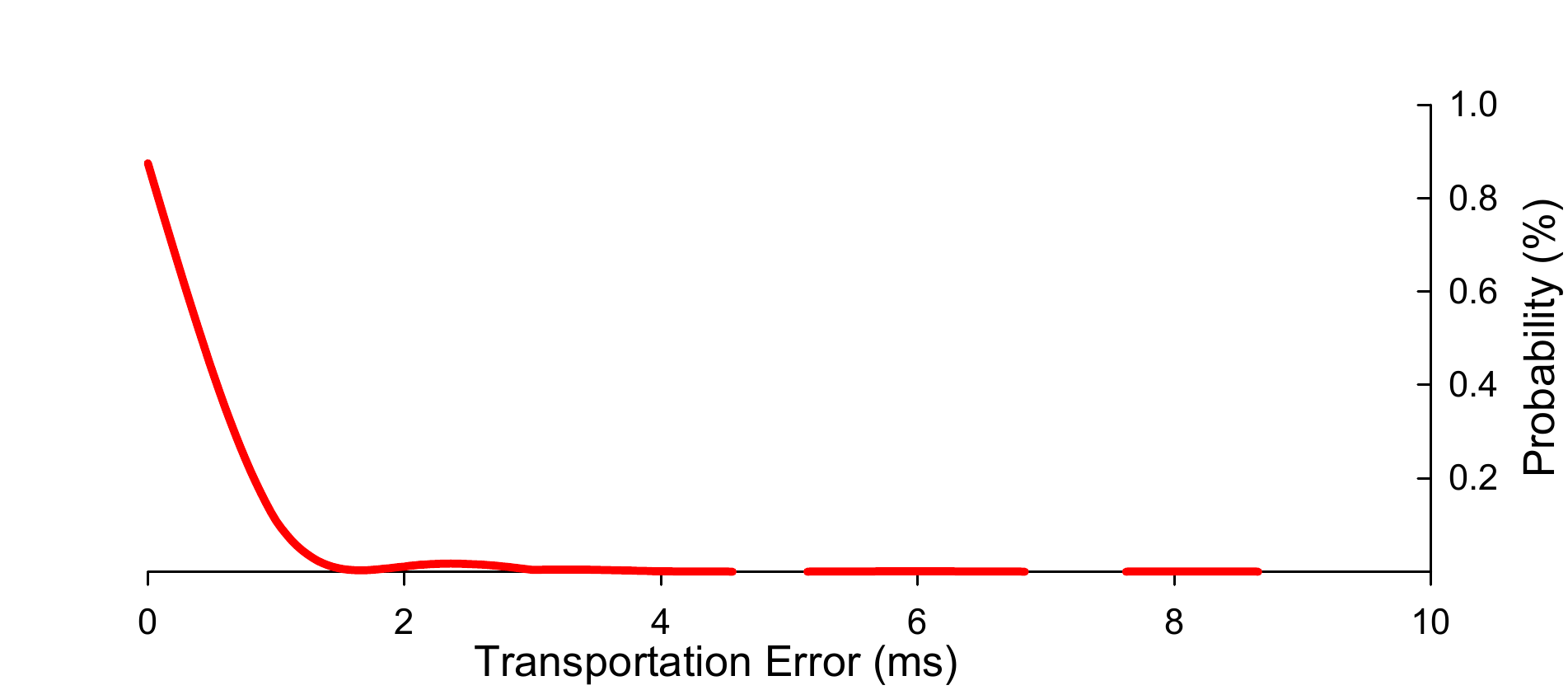}
				\par\end{centering}
		}
		\par\end{centering}
	\protect\caption{\label{fig:error_kinects} Results of the experiments on time synchronization. (a) Time error of each frame from Kinects $k_1$, $k_2$, and $k_3$ within 67.725 s. (b) Probability of error occurrence on the Kinect.
	}  
\end{figure}
The NTP {component is built }for synchronizing the clocks of the clients and the server, which guarantees that the designed time schedule mechanism commences its timing operations correctly. As mentioned in Section \ref{sc:Time Scheduling}, there should be 45 ms between each set of two continuous frames received by the server from the same client, which is defined as a time difference here. However, in practice, the time difference cannot be exactly 45 ms, as various factors might affect the transmission (such as networking conditions) {\cite{network_1}}.  We calculate the time differences and further compare them with the designed time cost (i.e., 45 ms) for evaluating the performance of the synchronization mechanism. The time error of a frame is defined as the difference between the actual time interval from such a frame to the previous one and the length of the designed time period. In a time period, each client occupies 15 ms as the time window for transformation. 
Therefore, if the time error of a frame is greater than 15 ms, the sequence in which the data are sent may be broken. 

\figurename \ref{fig:error_kinects_1} shows the transportation time error of each frame. The maximum values of the time errors of Kinect clients $K_1$, $K_2$, and $K_3$ in the experiment {are} 8 ms, 1 ms, and 4 ms, respectively, which are all less than the tolerable transportation error limit. The average transportation delay in the system is 1 ms. When the server wants to write the joint position data into the temporary file, such a thread would occupy the computing resource and lead to larger delays of frames from 2 ms to 7 ms.

\figurename \ref{fig:error_kinects_2} shows the probability of different degrees of transportation time error occurrences among the three Kinect sensors. The length of the experiment is 67.725 s, which includes the total 4518 frames received by the server. However, there are 74 frames that have transportation time errors greater than 1 ms, which means that the errors of 98.37\% of the frames are less than or equal to 1 ms.

{
\section {Accuracy Analysis}

\begin{figure}[!t]
	\begin{centering}
		\includegraphics[width=9cm]{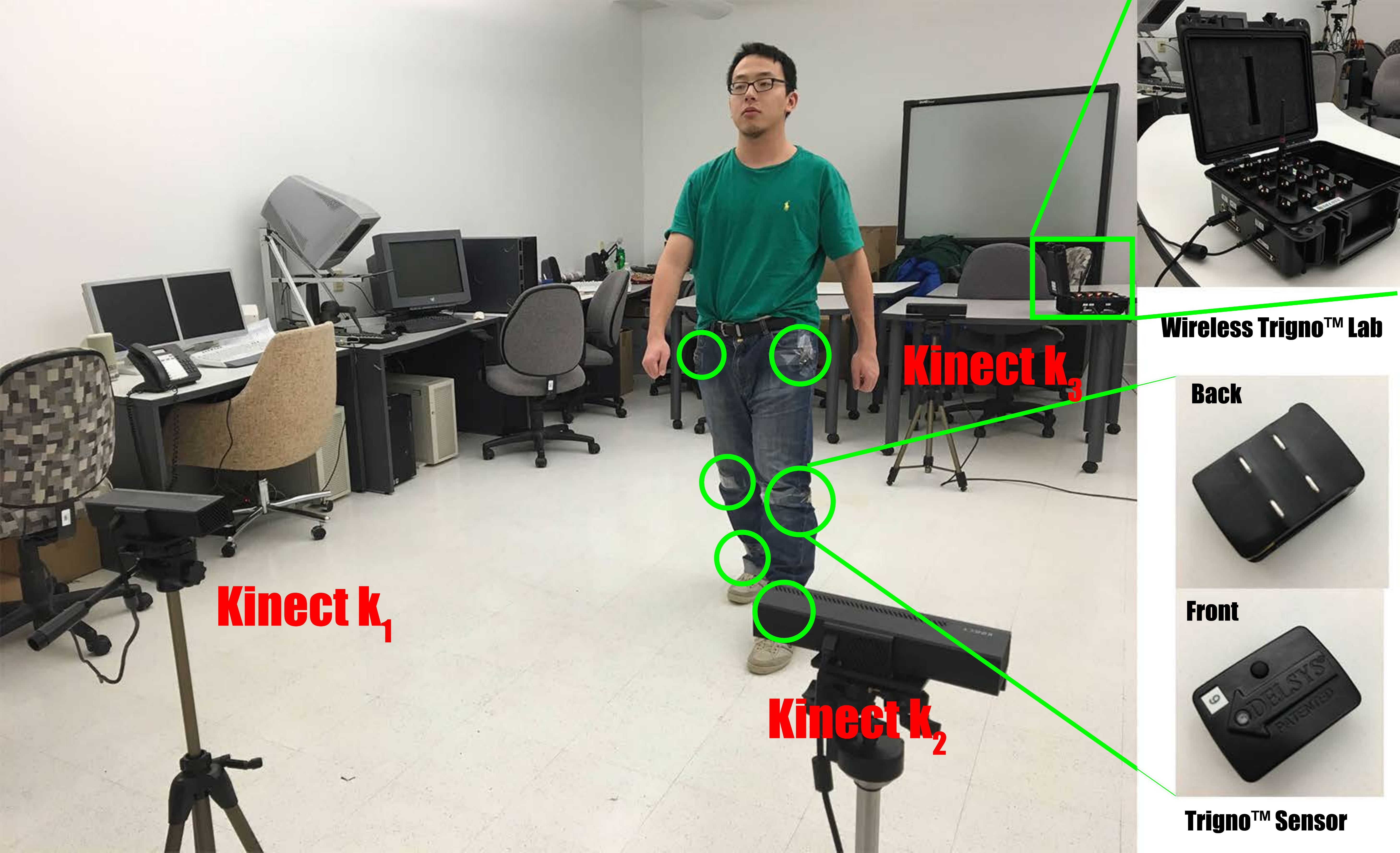}
		\par\end{centering}
	\protect\caption{\label{fig:comparative_analysis_photo}Accuracy comparision. Two motion capture systems (our proposed gait tracking system and the Delsys Trigno{\textsuperscript{TM}} Smart Sensor System) are compared to evaluate the accuracy of our system. Green circles indicate the positions of the tracked joints in human gait movement for both systems.}
\end{figure} 

Delsys Trigno{\textsuperscript{TM}} Wireless Smart Sensor System{\cite{trigno_1}} is chosen as the benchmark for evaluating the accuracy of our proposed system. Trigno{\textsuperscript{TM}} System is a kind of inertial-based motion capture system where each sensor has a built-in triaxial accelerometer, which can perceive a change in the sensor's acceleration and output it with the change in voltage.

The whole Trigno{\textsuperscript{TM}} System costs about \$1200, and this includes the software package, the server, and 16-channel sensors.} {In contrast, our system only costs about half of that amount due to the low price of the Kinect v2 sensors.
}

{
To evaluate the accuracy of our tracking system, we place 6 Trigno{\textsuperscript{TM}} smart sensors on the 6 joints of the lower limbs, including the left hip, left knee, left ankle, right hip, right knee, and right ankle, as shown in \figurename {\ref{fig:comparative_analysis_photo}}. Correspondingly, these six joints are measured by our proposed system simultaneously. In the experiment, a 2.5 s gait movement is measured using both of the two systems and compared. The tester is a 24-year-old male subject with a height of 176 cm and weight of 79 kg who moves his left and right feet forward and back one by one. The average distance between all the joints and Kinects is 346 cm as measured by a laser distance meter (distance measurement precision: $\pm$2 mm). As the Trigno{\textsuperscript{TM}} System is based on an inertial mechanism, meaning that the motion data streamed from the sensors are acceleration rather than position data, the acceleration data have to be transformed into position data before comparison.
}

\begin{figure*}[!t]
	\begin{centering}
		\includegraphics[width=19cm]{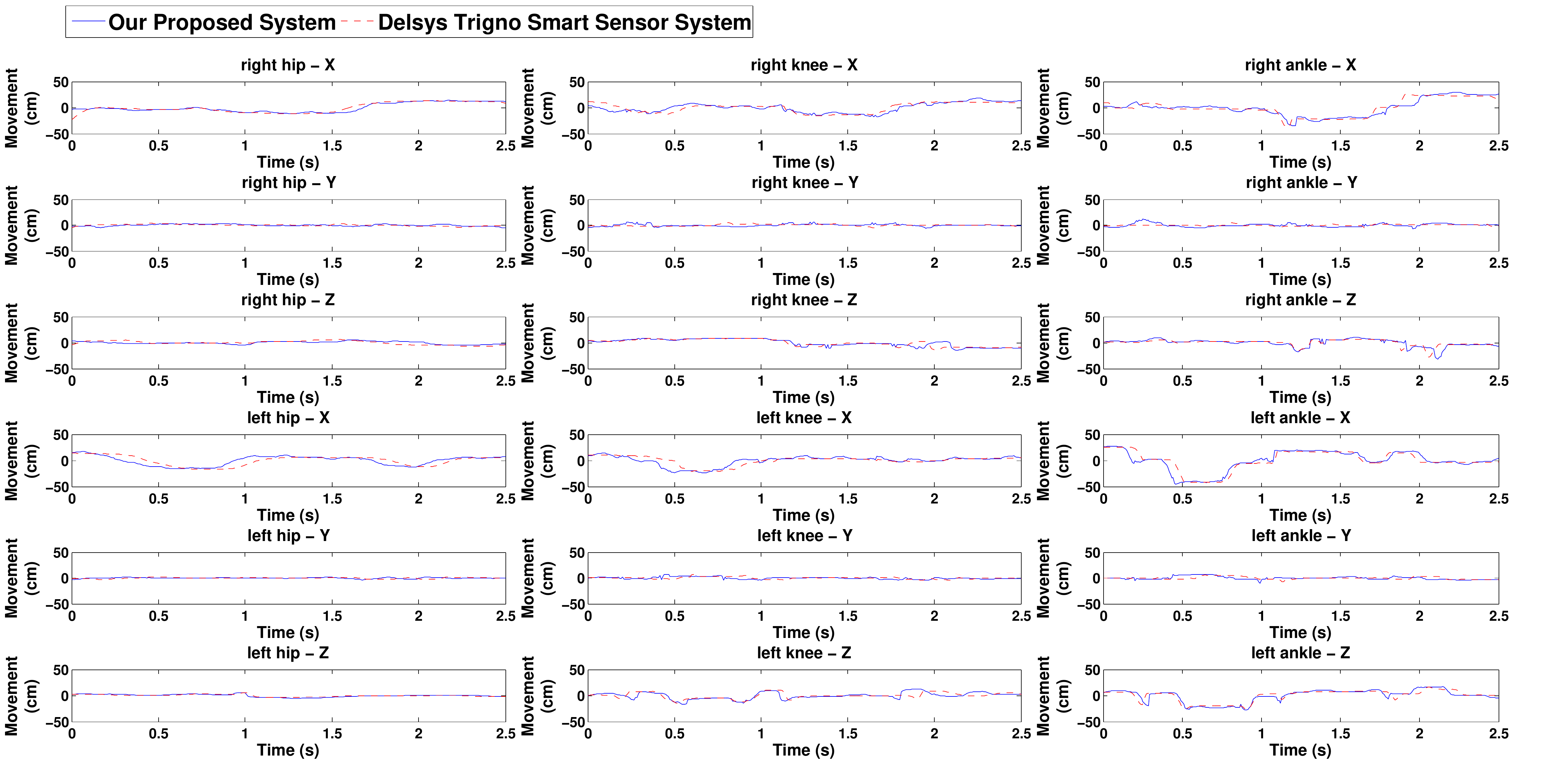}
		\par\end{centering}
	\protect\caption{\label{fig:comparative_analysis}Accuracy comparison results. The measured 3D coordinates of 6 joints (left hip, left knee, left ankle, right hip, right knee, right ankle) using both our proposed gait tracking system and Delsys Trigno{\textsuperscript{TM}} Wireless Smart Sensor System are compared. The blue solid line shows the trilateration measurement of  our proposed system; the green dashed line shows the inertia measurement of the Delsys System.}
\end{figure*} 

{
As shown in \figurename \ref{fig:comparative_analysis}, the two systems obtain the 6 joint positions in a total time interval of 2.5 s and report the triaxial results  after standardization and transformation. In the first 0.6 s, the tester moves his left foot  forward and moves it back in the next 0.6 s. During this process, it is shown that the largest changes occur in the left ankle's x-axis and z-axis positions. The y-axis position of the left ankle exhibits tiny changes because the direction of the left leg's movement is perpendicular to the y-axis of the coordinate system. Compared with the other two joints (i.e., left knee and left hip), a larger movement is observed from the ankle joint as compared with that from the hip and knee joints. Similarly, the movement of the right lower limb is measured in the time interval between 1.2 s and 2.5 s, which shows the same motion characteristics as those of the left leg.
}
	\begin{table}
		\begin{center}
			\protect\caption{\label{tab:system_comparison}Statistics of the Standard Deviation of the Difference between the Two Motion Capture Systems in  \figurename\ref{fig:comparative_analysis}}
			{
			\begin{center}
			\begin{tabular}{|>{\centering}p{2.5cm}|>{\centering}p{2.5cm}|>{\centering}p{5cm}|}
				\hline 
				{\scriptsize{}Coordinate} & {\scriptsize{}Standard Deviation of the Difference (cm)}\tabularnewline
				\hline 
				\hline 
				{\scriptsize{}Right hip - X axis}  & 
				{\scriptsize{}5.26}\tabularnewline 
				\hline
				{\scriptsize{}Right hip - Y axis}  & 
				{\scriptsize{}1.28}\tabularnewline 
				\hline
				{\scriptsize{}Right hip - Z axis}  & 
				{\scriptsize{}1.87}\tabularnewline 
				\hline
				{\scriptsize{}Right knee - X axis}  & 
				{\scriptsize{}4.48}\tabularnewline 
				\hline
				{\scriptsize{}Right knee - Y axis}  & 
				{\scriptsize{}1.16}\tabularnewline 
				\hline
				{\scriptsize{}Right knee - Z axis}  & 
				{\scriptsize{}1.19}\tabularnewline 
				\hline
				{\scriptsize{}Right ankle - X axis}  & 
				{\scriptsize{}3.58}\tabularnewline 
				\hline
				{\scriptsize{}Right ankle - Y axis}  & 
				{\scriptsize{}2.33}\tabularnewline 
				\hline
				{\scriptsize{}Right ankle - Z axis}  & 
				{\scriptsize{}3.40}\tabularnewline 
				\hline
				{\scriptsize{}Left hip - X axis}  & 
				{\scriptsize{}4.12}\tabularnewline 
				\hline
				{\scriptsize{}Left hip - Y axis}  & 
				{\scriptsize{}1.87}\tabularnewline 
				\hline
				{\scriptsize{}Left hip - Z axis}  & 
				{\scriptsize{}3.89}\tabularnewline 
				\hline
				{\scriptsize{}Left knee - X axis}  & 
				{\scriptsize{}6.90}\tabularnewline 
				\hline
				{\scriptsize{}Left knee - Y axis}  & 
				{\scriptsize{}2.60}\tabularnewline 
				\hline
				{\scriptsize{}Left knee - Z axis}  & 
				{\scriptsize{}4.37}\tabularnewline 
				\hline
				{\scriptsize{}Left ankle - X axis}  & 
				{\scriptsize{}5.81}\tabularnewline 
				\hline
				{\scriptsize{}Left ankle - Y axis}  & 
				{\scriptsize{}2.27}\tabularnewline 
				\hline
				{\scriptsize{}Left ankle - Z axis}  & 
				{\scriptsize{}4.24}\tabularnewline 
				\hline
				{\scriptsize{}Total}  & 
				{\scriptsize{}3.98}\tabularnewline 
				\hline

			\end{tabular}
		\end{center}
	}
		\end{center}
	\end{table}
{

We record the measurement from the Trigno{\textsuperscript{TM}} System as the benchmark and quantitatively provide the accuracy evaluation of our proposed system by computing the standard deviation of the measurement difference between the two motion capture systems (shown in Table \ref{tab:system_comparison}). Based on the comparison of the different axes of the same joint, the maximum errors always} {occur at the x axis, along which joints make the maximum movements. On the other hand, the technical specification of the Trigno{\textsuperscript{TM}} System shows that the standard deviation of the voltage output error of the sensors is 0.233 V (i.e., 0.94 cm after transformation). Furthermore, we obtain the standard deviation of the measurement error of our proposed system by adding the total standard deviation of the measurement difference between the two motion systems as 4.92 cm=0.94 cm+3.98 cm under the circumstance that the average distance from all joints to the Kinect sensors is 346 cm.
}

\section{Conclusion}

Based on the proposed geometric trilateration method, six joints on the user's legs are located by our human gait tracking system using three Kinect v2 sensors. During this process, the proposed synchronization and time scheduling mechanisms allow the system to receive and collect data accurately when the data transmission is affected by various factors such as the network conditions. The experimental result shows that there are a few frames with small time delays, but none of them cause incorrect trilateration results. { The accuracy of our proposed system is evaluated and verified by comparison with a commercial medical system (Delsys Trigno Smart Sensor System).}


%





\ifCLASSOPTIONcaptionsoff
  \newpage
\fi



%

\bibliographystyle{ieeetr}
\bibliography{Bibliov}

\begin{thebibliography}{10}

\bibitem{intro_mc_1}
G.~Johansson, ``Visual perception of biological motion and a model for its
  analysis,'' {\em Attention, Perception, \& Psychophysics}, vol.~14, no.~2,
  pp.~201--211, 1973.

\bibitem{intro_mc_4}
C.~M. Ginsberg and D.~Maxwell, ``Graphical marionette,'' in {\em Proceeding of
  the ACM SIGGRAPH/SIGART Interdisciplinary Workshop on Motion: Representation
  and Perception}, pp.~303--310, 1986.

\bibitem{intro_mc_5}
J.~Sabel, ``Optical 3{D} motion measurement,'' in {\em Proceedings of
  Instrumentation and Measurement Technology Conference on Indispensable Bridge
  between Theory and Reality}, vol.~1, pp.~367--370, 1996.

\bibitem{intro_mc_10}
C.~Bregler and J.~Malik, ``Tracking people with twists and exponential maps,''
  in {\em Proceedings of IEEE Computer Society Conference on Computer Vision
  and Pattern Recognition}, pp.~8--15, 1998.

\bibitem{intro_mc_7}
B.~M. Thomas and E.~Granum, ``A survey of computer vision-based human motion
  capture,'' {\em Computer Vision and Image Understanding}, vol.~81, no.~3,
  pp.~231 -- 268, 2001.

\bibitem{intro_mc_8}
T.~Xing, Y.~Yu, Y.~Zhou, and S.~Du, ``Markerless motion capture of human body
  using {PSO} with single depth camera,'' in {\em Proceedings of Second
  International Conference on 3{D} Imaging, Modeling, Processing, Visualization
  and Transmission}, pp.~192--197, 2012.

\bibitem{intro_mc_9}
T.~B. Moeslund, A.~Hilton, and V.~Kr{\"u}ger, ``A survey of advances in
  vision-based human motion capture and analysis,'' {\em Computer Vision and
  Image Understanding}, vol.~104, no.~2, pp.~90 -- 126, 2006.

\bibitem{intro_mc_11}
D.~Reber, ``Nick strives to define motion capture,'' {\em Animation World
  Magzine}, vol.~3, p.~11, 1999.

\bibitem{intro_mc_16}
A.~Lees and L.~Nolan, ``The biomechanics of soccer: {A} review,'' {\em Journal
  of Sports Sciences}, vol.~16, no.~3, pp.~211--234, 1998.

\bibitem{intro_mc_17}
J.~L. Hudson, ``A biomechanical analysis by skill level of free throw shooting
  in basketball,'' {\em Biomechanics in Sports}, pp.~95--102, 1982.

\bibitem{intro_mc_18}
T.~Nikodelis, I.~Kollias, and V.~Hatzitaki, ``Bilateral inter-arm coordination
  in freestyle swimming: Effect of skill level and swimming speed,'' {\em
  Journal of Sports Sciences}, vol.~23, no.~7, pp.~737--745, 2005.

\bibitem{intro_mc_14}
``Vicon motion systems.'' http://www.vicon.com/, Accessed April 2, 2015.

\bibitem{Intro_mc_19}
S.~Dent, ``What you need to know about 3{D} motion capture.''
  http://www.engadget.com/2014/07/14/motion-capture-explainer/, Accessed July
  14, 2015.

\bibitem{Zhang_2015}
L.~Zhang, H.~Dong, and A.~El~Saddik, ``From 3{D} sensing to printing: A
  survey,'' {\em ACM Transactions on Multimedia Computing, Communications and
  Applications}, vol.~12, no.~2, pp.~27:1--23, 2015.

\bibitem{Figueroa_2015}
N.~Figueroa, H.~Dong, and A.~El~Saddik, ``A combined approach towards
  consistent reconstructions of indoor spaces based on 6{D} {RGB-D} odometry
  and {KinectFusion},'' {\em ACM Transactions on Intelligent Systems and
  Technology}, vol.~6, no.~2, pp.~14:1--10, 2015.

\bibitem{SO_1}
L.~Yang, L.~Zhang, H.~Dong, A.~Alelaiwi, and A.~El~Saddik, ``Evaluating and
  improving the depth accuracy of {K}inect for {W}indows v2,'' {\em IEEE
  Sensors Journal}, vol.~15, no.~8, pp.~4275--4285, 2015.

\bibitem{related_mc_1}
S.~L. Dockstader and A.~M. Tekalp, ``Multiple camera tracking of interacting
  and occluded human motion,'' {\em Proceedings of the IEEE}, vol.~89, no.~10,
  pp.~1441--1455, 2001.

\bibitem{RL_7}
Y.~Han, ``2{D}-to-3{D} visual human motion converting system for home optical
  motion capture tool and 3-{D} smart {TV},'' {\em IEEE Systems Journal},
  vol.~9, no.~1, pp.~131--140, 2015.

\bibitem{related_mc_2}
T.~Cloete and C.~Scheffer, ``Benchmarking of a full-body inertial motion
  capture system for clinical gait analysis,'' in {\em Proceedings of 30th
  Annual International Conference on Engineering in Medicine and Biology
  Society}, pp.~4579--4582, 2008.

\bibitem{related_mc_4}
E.~H. Hall, ``On a new action of the magnet on electric currents,'' {\em
  American Journal of Mathematics}, vol.~2, no.~3, pp.~287--292, 1879.

\bibitem{related_mc_6}
A.~Gmiterko and T.~Lipt{\'a}k, ``Motion capture of human for interaction with
  service robot,'' {\em American Journal of Mechanical Engineering}, vol.~1,
  no.~7, pp.~212--216, 2013.

\bibitem{related_mc_7}
J.~Vince, {\em Essential Computer Animation fast: How to Understand the
  Techniques and Potential of Computer Animation}.
\newblock 2000.

\bibitem{RL_3}
D.~Weinland, R.~Ronfard, and E.~Boyer, ``A survey of vision-based methods for
  action representation, segmentation and recognition,'' {\em Computer Vision
  and Image Understanding}, vol.~115, no.~2, pp.~224--241, 2011.

\bibitem{RL_4}
M.~Ye, Q.~Zhang, L.~Wang, J.~Zhu, R.~Yang, and J.~Gall, ``A survey on human
  motion analysis from depth data,'' in {\em Time-of-Flight and Depth Imaging.
  Sensors, Algorithms, and Applications}, vol.~8200, pp.~149--187, Springer
  Berlin Heidelberg, 2013.

\bibitem{RL_5}
L.~Chen, H.~Wei, and J.~Ferryman, ``A survey of human motion analysis using
  depth imagery,'' {\em Pattern Recognition Letters}, vol.~34, no.~15,
  pp.~1995--2006, 2013.

\bibitem{RL_6}
J.~Aggarwal and L.~Xia, ``Human activity recognition from 3{D} data: {A}
  review,'' {\em Pattern Recognition Letters}, vol.~48, pp.~70--80, 2014.

\bibitem{001KinectWeb}
``Kinect.'' http://www.microsoft.com/en-us/kinectforwindows/, Accessed August
  28, 2014.

\bibitem{011Parts}
J.~Shotton, T.~Sharp, A.~Kipman, A.~Fitzgibbon, M.~Finocchio, A.~Blake,
  M.~Cook, and R.~Moore, ``Real-time human pose recognition in parts from
  single depth image,'' {\em Communications of the ACM}, vol.~56, no.~1,
  pp.~116--124, 2013.

\bibitem{RL_8}
M.~Gupta, L.~Behera, V.~K. Subramanian, and M.~M. Jamshidi, ``A robust visual
  human detection approach with {UKF}-based motion tracking for a mobile
  robot,'' {\em IEEE Systems Journal}, 2014.

\bibitem{RL_gait_2}
B.~Galna, G.~Barry, D.~Jackson, D.~Mhiripiri, P.~Olivier, and L.~Rochester,
  ``Accuracy of the {M}icrosoft {K}inect sensor for measuring movement in
  people with {P}arkinson's disease,'' {\em Gait and Posture}, vol.~39, no.~4,
  pp.~1062--1068, 2014.

\bibitem{RL_gait_3}
X.~Xu, R.~W. McGorry, L.-S. Chou, J.-H. Lin, and C.-C. Chang, ``Accuracy of the
  microsoft {K}inect\textsuperscript{TM} for measuring gait parameters during
  treadmill walking,'' {\em Gait and Posture}, vol.~42, no.~2, pp.~145--151,
  2015.

\bibitem{RL_gait_4}
E.~Auvinet, F.~Multon, C.-E. Aubin, J.~Meunier, and M.~Raison, ``Detection of
  gait cycles in treadmill walking using a {K}inect,'' {\em Gait and Posture},
  vol.~41, no.~2, pp.~722--725, 2015.

\bibitem{RL_gait_5}
L.~F. Yeung, K.~C. Cheng, C.~H. Fong, W.~C.~C. Lee, and K.-Y. Tong,
  ``Evaluation of the {M}icrosoft {K}inect as a clinical assessment tool of
  body sway,'' {\em Gait and Posture}, vol.~40, no.~4, pp.~532--538, 2014.

\bibitem{RL_gait_6}
B.~F. Mentiplay, L.~G. Perraton, K.~J. Bower, Y.-H. Pua, R.~McGaw, S.~Heywood,
  and R.~A. Clark, ``Gait assessment using the {M}icrosoft {X}box {O}ne
  {K}inect: Concurrent validity and inter-day reliability of spatiotemporal and
  kinematic variables,'' {\em Journal of Biomechanics}, vol.~48, no.~10,
  pp.~2166--2170, 2015.

\bibitem{RL_gait_1}
R.~A. Clark, Y.-H. Pua, C.~C. Oliveira, K.~J. Bower, S.~Thilarajah, R.~McGaw,
  K.~Hasanki, and B.~F. Mentiplay, ``Reliability and concurrent validity of the
  {M}icrosoft {X}box {O}ne {K}inect for assessment of standing balance and
  postural control,'' {\em Gait and Posture}, vol.~42, no.~2, pp.~210--213,
  2015.

\bibitem{02time_sync}
L.~D. Mills, ``Network time protocol (version 3) specification.''
  Implementation and Analysis - RFC1305, 1992.

\bibitem{024determination}
W.~Hereman and S.~W. Murphy, ``Determination of a position in three dimensions
  using trilateration and approximate distances,'' tech. rep., Colorado School
  of Mines, 1995.

\bibitem{usb_1}
``The {USB}3.1 {S}pecification.'' http://www.usb.org/developers/, Accessed Feb
  9, 2016.

\bibitem{ieee_1588_1}
J.~Eidson, ``{IEEE} 1588 {S}tandard {V}ersion 2 - {A} {T}utorial.''
  http://www.webcitation.org/5qaJpYqCH, Accessed Feb 9, 2016.

\bibitem{gaitsys_ts_1}
S.~Wainner and R.~Richmond, {\em The Book of Overclocking: Tweak Your PC to
  Unleash Its Power}.
\newblock 2003.

\bibitem{network_1}
F.~Halsall and D.~Links, {\em Computer Networks and Open Systems}.
\newblock Addison-Wesley Publishers, 1995.

\bibitem{trigno_1}
``{T}echnical {S}pecification of {T}rigno{\textsuperscript{tm}} {L}ab.''
  http://www.delsys.com/products/wireless-emg/trigno-lab/, Accessed Feb 9,
  2016.

\end{thebibliography}


%
\begin{IEEEbiography}[{\includegraphics[width=1in,height=1.25in,clip,keepaspectratio]{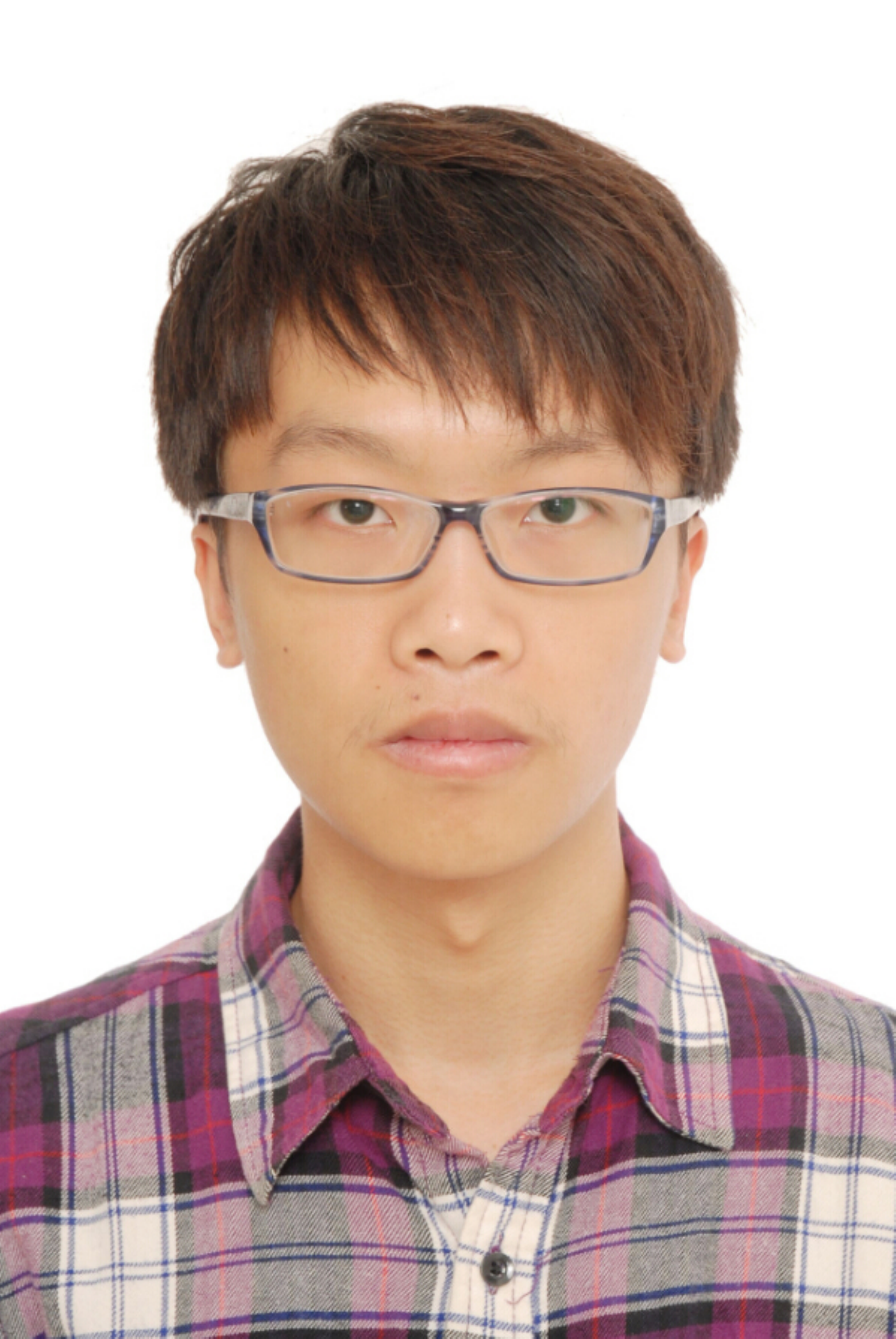}}]{Lin Yang}
is currently pursuing his Ph.D. degree in the National University of Singapore (NUS). He received his B.Eng. in Software Engineering from Sichuan University in 2013 and M.Sc. at the University of Ottawa in 2015. He had been working on the project named Touchable Avatar at MCRLab at the University of Ottawa. His research interests include computer graphics, computer vision and human-machine interaction.
\end{IEEEbiography}

\begin{IEEEbiography}[{\includegraphics[width=1in,height=1.25in,clip,keepaspectratio]{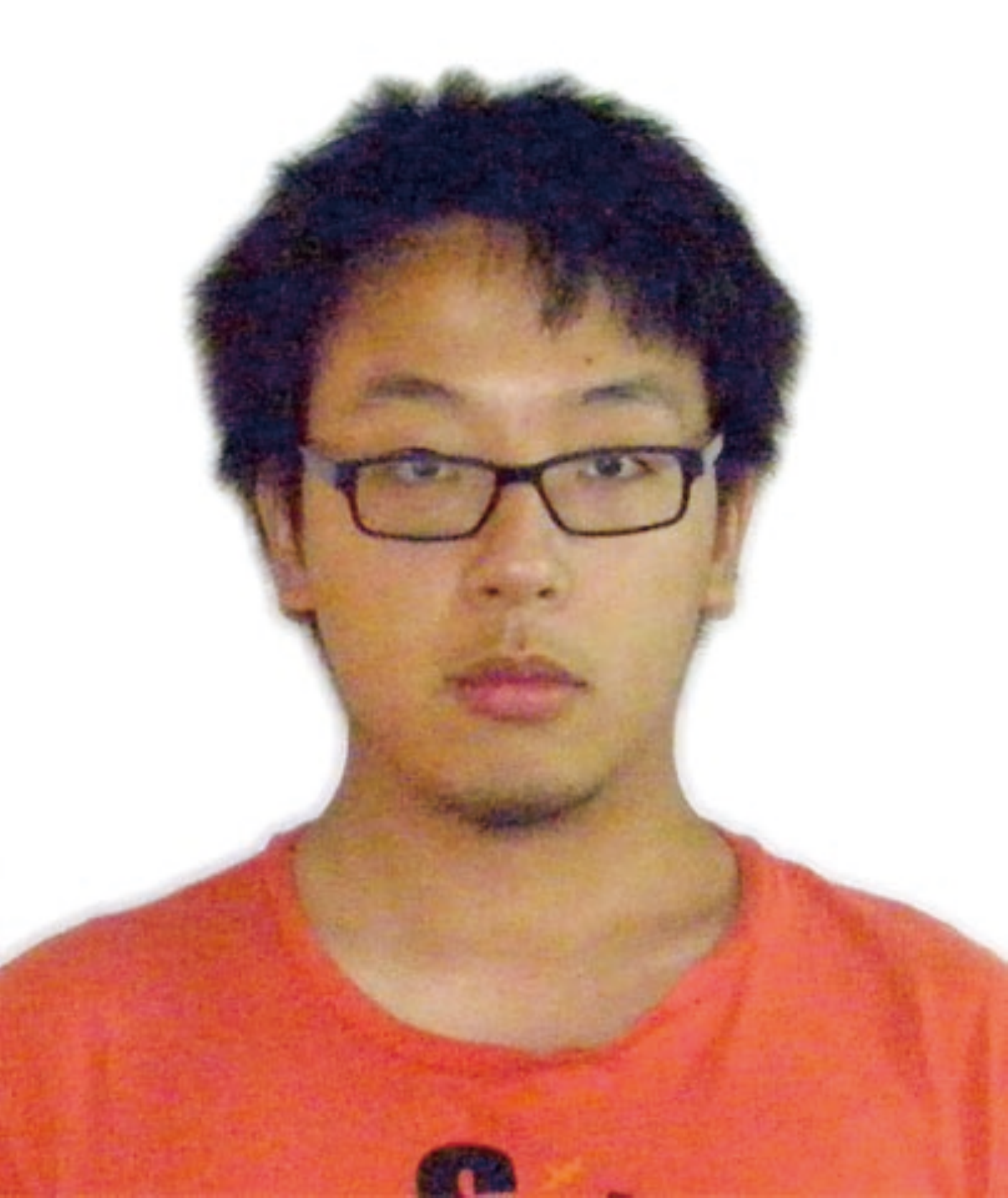}}]{Bowen Yang}
	received his B.Eng. in Computer Science from Central South University in 2014 and started his M.Sc. at the University of Ottawa in the same year. He is currently working on the project of 3D Sensing and Tracking of Human Gait Movement, which involves finding an optimal solution to improve the accuracy of multiple Microsoft Kinect v2 sensors for medical purposes. His research interests include computer vision, nonlinear optimization, and calibration strategy.
\end{IEEEbiography}


\begin{IEEEbiography}[{\includegraphics[width=1in,height=1.25in,clip,keepaspectratio]{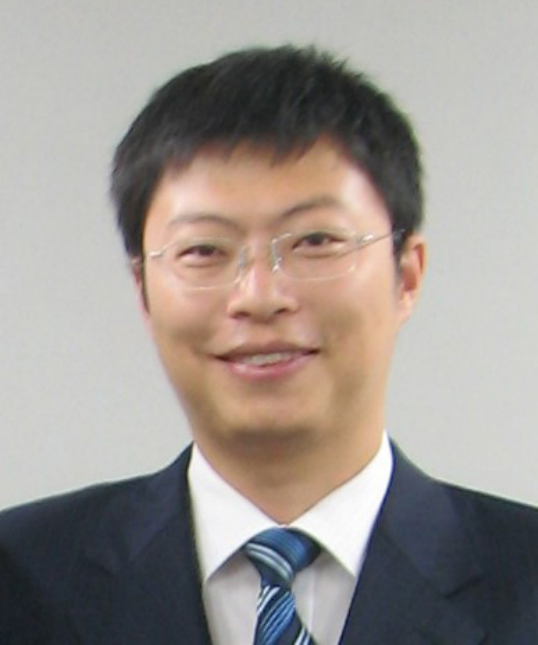}}]{Haiwei Dong (M’12 SM’16)}
received Dr.Eng. in Computer Science and Systems Engineering and M.Eng. in Control Theory and Control Engineering from Kobe University (Japan) and Shanghai Jiao Tong University (P.R.China) in 2010 and 2008, respectively. He is currently with the University of Ottawa. Prior to that, he was appointed as Postdoctoral Fellow at New York University Abu Dhabi, Research Associate at the University of Toronto, Research Fellow (PD) at the Japan Society for the Promotion of Science (JSPS), Science Technology Researcher at Kobe University, and Science Promotion Researcher at Kobe Biotechnology Research and Human Resource Development Center. His research interests include robotics, haptics, control and multimedia. He is a Senior Member of IEEE.
\end{IEEEbiography}


\begin{IEEEbiography}[{\includegraphics[width=1in,height=1.25in,clip,keepaspectratio]{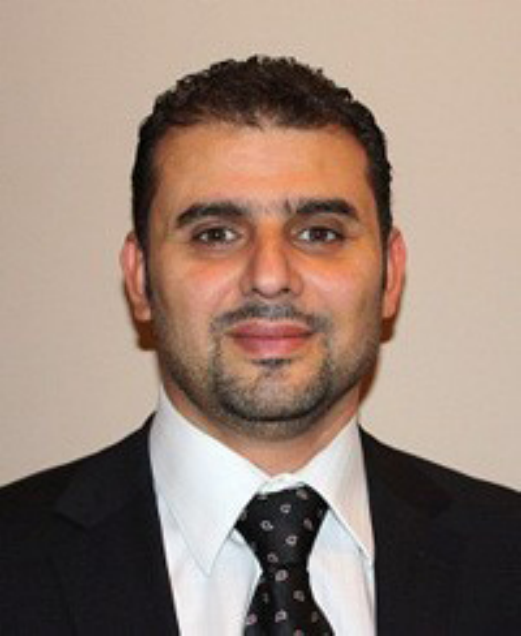}}]{Abdulmotaleb El Saddik (M’01 SM’04 F’09)}
is Distinguished University Professor and University Research Chair in the School of Electrical Engineering and Computer Science at the University of Ottawa. His research focus on multimodal interaction with multimedia information in smart cities. He is an internationally-recognized scholar who has made strong contributions to the knowledge and understanding of multimedia computing, communications and applications. He has authored and co-authored four books and more than 450 publications. Chaired more than 40 conferences and workshop and has received research grants and contracts totaling more than \$18 Mio. He has supervised more than 100 researchers. He received several international awards, among others are ACM Distinguished Scientist, Fellow of the Engineering Institute of Canada, Fellow of the Canadian Academy of Engineers and Fellow of IEEE and IEEE Canada Computer Medal.
\end{IEEEbiography}
\vfill




\end{document}